\documentclass[aps,prl,twocolumn,reprint,amsmath,amssymb,superscriptaddress,floatfix,footinbib,longbibliography]{revtex4-1}

\usepackage{epsfig}
\usepackage{graphicx}
\usepackage{epstopdf}

\usepackage[T1]{fontenc}
\usepackage[applemac]{inputenc}
\usepackage{lmodern}
\usepackage[english]{babel}

\usepackage{ae}
\usepackage{units}

\usepackage{amsmath,amssymb,natbib,bm}
\usepackage{psfrag}
\usepackage{subfigure}
\usepackage{amsthm}

\usepackage{color}
\usepackage{url}

\usepackage[colorlinks]{hyperref}
\hypersetup{%
        plainpages=true,
        breaklinks=true,
        hypertexnames=false,
        pageanchor=true,
        colorlinks=true,
        linkcolor={blue},
        citecolor={magenta},
        urlcolor={blue},
        anchorcolor={black}
      }
      
\usepackage{mleftright} 

\usepackage{lineno}
\modulolinenumbers[5]

\usepackage{framed}

\usepackage[x11names]{xcolor}
\colorlet{shadecolor}{LightSteelBlue1}
\colorlet{framecolor}{LightSteelBlue4}

\newenvironment{frshaded}{%
\MakeFramed {\FrameRestore}}%
{\endMakeFramed}

\newcommand{\figref}[1]{\mbox{Fig.~\ref{#1}}}

\renewcommand{\eqref}[1]{\mbox{Eq.~(\ref{#1})}}

\newcommand{\ket}[1]{|#1\rangle}

\newcommand{\ketbra}[2]{\mleft| #1 \rangle \langle #2 \mright|}
\newcommand{\brakket}[3]{\langle #1 | #2 | #3 \rangle}
\newcommand{\expec}[1]{\mleft\langle #1 \mright\rangle}

\newcommand{\sz}{\hat \sigma_z}
\newcommand{\sx}{\hat \sigma_x}
\newcommand{\sm}{\hat \sigma_-}
\renewcommand{\sp}{\hat \sigma_+}

\newcommand{\abssq}[1]{\mleft| #1 \mright|^2}

\newcommand{\be}{\begin{equation}}
\newcommand{\ee}{\end{equation}}
\newcommand{\bea}{\begin{eqnarray}}
\newcommand{\eea}{\end{eqnarray}}

\begin{document}

\title{Ultrastrong coupling between light and matter}

\author{Anton Frisk Kockum}
\email[e-mail:]{anton.frisk.kockum@gmail.com}
\affiliation{Theoretical Quantum Physics Laboratory, RIKEN Cluster for Pioneering Research, Wako-shi, Saitama 351-0198, Japan}

\author{Adam Miranowicz}
\affiliation{Faculty of Physics, Adam Mickiewicz University, 61-614 Pozna\'n, Poland}
\affiliation{Theoretical Quantum Physics Laboratory, RIKEN Cluster for Pioneering Research, Wako-shi, Saitama 351-0198, Japan}

\author{Simone De Liberato}
\affiliation{School of Physics and Astronomy, University of Southampton, Southampton, SO17 1BJ, United Kingdom}
\affiliation{Theoretical Quantum Physics Laboratory, RIKEN Cluster for Pioneering Research, Wako-shi, Saitama 351-0198, Japan}

\author{Salvatore Savasta}
\affiliation{Dipartimento di Scienze Matematiche e Informatiche, Scienze Fisiche e Scienze della Terra, Universit\`a di Messina, I-98166 Messina, Italy}
\affiliation{Theoretical Quantum Physics Laboratory, RIKEN Cluster for Pioneering Research, Wako-shi, Saitama 351-0198, Japan}

\author{Franco Nori}
\email[e-mail:]{fnori@riken.jp}
\affiliation{Theoretical Quantum Physics Laboratory, RIKEN Cluster for Pioneering Research, Wako-shi, Saitama 351-0198, Japan}
\affiliation{Physics Department, The University of Michigan, Ann Arbor, Michigan 48109-1040, USA}

\date{submitted May 7, revised July 30}

\begin{abstract}

\textbf{
Ultrastrong coupling between light and matter has, in the past decade, transitioned from theoretical idea to experimental reality. It is a new regime of quantum light-matter interaction, going beyond weak and strong coupling to make the coupling strength comparable to the transition frequencies in the system. The achievement of weak and strong coupling has led to increased control of quantum systems and applications like lasers, quantum sensing, and quantum information processing. Here we review the theory of quantum systems with ultrastrong coupling, which includes entangled ground states with virtual excitations, new avenues for nonlinear optics, and connections to several important physical models. We also review the multitude of experimental setups, including superconducting circuits, organic molecules, semiconductor polaritons, and optomechanics, that now have achieved ultrastrong coupling. We then discuss the many potential applications that these achievements enable in physics and chemistry.
}

\end{abstract}

\maketitle


\section{Introduction}

The intuitive description of the interaction between light and matter as a series of elementary processes, in which a photon is absorbed, emitted, or scattered by a distribution of charges, essentially hinges on the small value of the fine structure constant $\alpha \simeq \frac{1}{137}$. Being $\alpha$ the natural dimensionless parameter emerging in a perturbative treatment of quantum electrodynamics, its small value allows to describe most of the quantum dynamics of the electromagnetic field by only taking into account first-order (absorption, emission) or second-order (scattering) processes.

\begin{figure*}
\centering
\includegraphics[width=\linewidth]{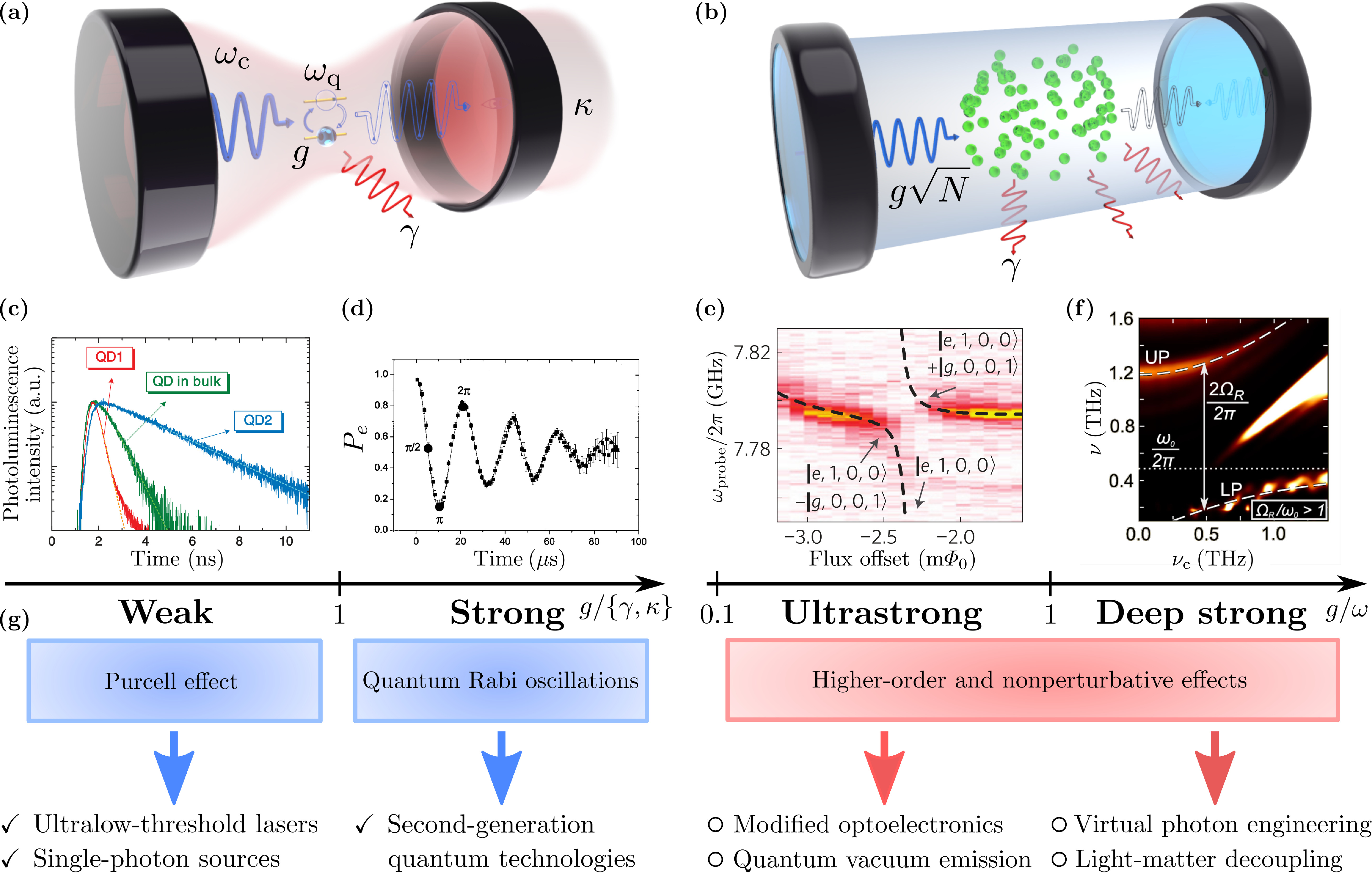}
\caption{Regimes of light-matter interaction.
\textbf{(a)} Sketch of a CQED system with a single two-level atom (qubit; the simplest example of a matter excitation). The parameters determining the different interaction regimes are the resonance frequency $\omega_{\rm c}$ of the cavity mode, the transition frequency $\omega_{\rm q}$ of the qubit, the coupling strength $g$, and the cavity and qubit loss rates: $\kappa$ and $\gamma$.
\textbf{(b)} Sketch of an optical resonator coupled to many quantum emitters. The light-matter coupling strength can be enhanced by increasing the number $N$ of emitters interacting with the resonator. The resulting collective coupling strength scales as $g \sqrt{N}$, where $g$ is the coupling between the light and a single emitter. 
\textbf{(c)-(f)} Four representative CQED experiments illustrating different light-matter interaction regimes.
\textbf{(c)} \textbf{Weak coupling}: Experimental demonstration of full control of the spontaneous-emission dynamics of single quantum dots (QDs) by a photonic-crystal nanocavity~\cite{Chang2006}. The plot shows time-resolved micro-photoluminescence intensities of InGaAs quantum dots on resonance with the cavity (QD1), off resonance (QD2), and in bulk without any cavity. Compared to the case without any cavity, the QDs decay faster in the presence of a resonant cavity (which enhances the density of states that the QDs can decay to) and slower in the presence of an off-resonant cavity (which shields the QD from the environment). This is the Purcell effect~\cite{Purcell1946}.
\textbf{(d)} \textbf{Strong coupling}: Data from a pioneering experiment~\cite{Raimond2001} with Rydberg atoms coupled to a superconducting microwave Fabry-Perot resonator, displaying vacuum Rabi oscillations. An atom in the excited state $\ket{e}$ enters an empty resonant cavity and the excitation is exchanged back and forth between the atom and the resonator before it decays. $P_e$ denotes the probability of detecting the atom in $\ket{e}$ as a function of the effective interaction time.
\textbf{(e)} \textbf{Ultrastrong coupling}: Microwave spectroscopy of a system with a superconducting flux qubit coupled to a coplanar-waveguide resonator~\cite{Niemczyk2010}. The system displays a normalized coupling strength $\eta= g/ \omega_{\rm c} = 0.12$. The plot shows the cavity transmission as a function of probe frequency $\omega_{\rm probe}$ and flux offset, which tunes the qubit frequency. The avoided level crossing indicates a coupling between states with different numbers of excitations (one state has a single photon in the third resonator mode; the other state has one qubit excitation and one photon in the first resonator mode). Such a coupling requires counter-rotating terms and is not reproduced by the Jaynes-Cummings approximation (see Box~1 and the adjacent discussion).
\textbf{(f)} \textbf{Deep strong coupling}: Magneto-THz transmission measurements on a THz metamaterial coupled to the cyclotron resonance of a two-dimensional electron gas~\cite{Bayer2017}. The splitting $2\Omega_R$ between the lower polariton (LP) and upper polariton (UP) levels that emerges as the cyclotron frequency $\nu_c$ is tuned, is a measure of the coupling strength. In this work, a record normalized coupling of $\eta = 1.43$ was reached.
\textbf{(g)} Phenomena and applications associated with different strengths of light-matter interaction.
Figures reproduced with permission from: \textbf{(c)} Ref.~\cite{Chang2006} \copyright 2006, APS; \textbf{(d)} Ref.~\cite{Raimond2001}, \copyright 2001, APS; \textbf{(e)} Ref.~\cite{Niemczyk2010} \copyright 2010, NPG; and \textbf{(f)} Ref.~\cite{Bayer2017} \copyright 2017, ACS.
\label{fig:SetupAndRegimes}}
\end{figure*}

While the value of $\alpha$ is fixed by nature, Purcell discovered in 1946 that the intensity of the interaction of an emitter with light can be enhanced or suppressed by engineering its electromagnetic environment~\cite{Purcell1946}. From this crucial observation sprung a whole field of research, today called cavity quantum electrodynamics [CQED, see \figref{fig:SetupAndRegimes}(a)], which aims to exploit different kinds of photonic resonators to modulate the coupling of light with matter.

The fundamental and applied importance of controlling the strength of light-matter coupling $g$ led to the development of resonators with ever higher quality factors. In 1983, Haroche and co-workers, using a collection of Rydberg atoms in a high-$Q$ microwave cavity, managed to achieve a coupling strength exceeding the losses in the system~\cite{Kaluzny1983}. In this strong-coupling [SC, see \figref{fig:SetupAndRegimes}(d)] regime, it is possible to observe an oscillatory exchange of energy quanta between the matter and the light, called vacuum Rabi oscillations, which takes place at a rate given by $g$. In the weak-coupling [WC, see \figref{fig:SetupAndRegimes}(c)] regime, when $g$ is smaller than the losses, the energy is instead lost from the system before it can be exchanged between the light and the matter.  

The SC regime was soon also reached with single atoms coherently interacting with a microwave cavity~\cite{Meschede1985} and, a few years later, with an optical cavity~\cite{Thompson1992}. In 1992, the SC regime was demonstrated using quasi-2D electronic excitations (Wannier excitons) embedded in a semiconductor optical microcavity~\cite{Weisbuch1992}. In this case, the resulting system eigenstates are called cavity-polaritons. Following these pioneering experiments, CQED has been successfully adapted and further developed using artificial atoms, such as quantum dots~\cite{Lodahl2015} and superconducting qubits (circuit QED)~\cite{Gu2017}.

In a CQED setup, the dimensionless parameter quantifying the interaction is the ratio between the coupling strength $g$ and the bare energy of the excitations. This quantity, the normalized coupling $\eta$, is proportional to a positive power of $\alpha$ and its value in the first observations of the SC regime was smaller than $10^{-6}$ for atoms \cite{Thompson1992} and $10^{-3}$ for Wannier excitons in semiconductor microcavities~\cite{Weisbuch1992}. Lowest-order perturbation theory is thus perfectly adequate to describe those experiments. The important difference with the WC regime is that, being the coupling larger than the spectral width of the excitations, degenerate perturbation theory needs to be applied.

It took more than two decades after the observation of SC for the CQED community to begin investigating the possibility to access a regime with larger $\eta$ in which higher-order processes, which would hybridize states with different number of excitations, become observable. Two main paths were identified to reach such a regime. The first is to couple many dipoles to the same cavity mode [\figref{fig:SetupAndRegimes}(b)]. As correctly predicted by the Dicke model~\cite{Dicke1954}, this leads to an enhanced coupling which scales with the square root of the number of dipoles. The second path is to use different degrees of freedom, whose coupling is not bounded by the small value of $\alpha$~\cite{Devoret2007}.

In 2005, following the first path, it was predicted~\cite{Ciuti2005} that this regime, which was named the ultrastrong-coupling [USC, see \figref{fig:SetupAndRegimes}(e)] regime, could be observed in intersubband polaritons thanks to the large number of electrons involved in the transitions between parallel subbands in a quantum well. In 2009, the USC regime was effectively observed for the first time in a microcavity-embedded doped GaAs quantum well, with $\eta=0.11$~\cite{Anappara2009}. Following this initial observation, the value of $\eta=0.1$ has often been taken as a threshold for the USC regime. While useful, it is important to note that the intensity of higher-order processes depend continuously on $\eta$, and the value of $0.1$ is thus just a historical convention, without any deeper physical meaning.

The second path has been followed in experiments with superconducting circuits~\cite{Gu2017}, in which ultrastrong coupling was observed in 2010, with $\eta=0.12$~\cite{Niemczyk2010}. In these experiments, it becomes possible to explore USC of light to a single two-level system, instead of a collective excitation.

Following these experimental breakthroughs, the interest in USC has blossomed, fostered by the vast phenomenology which has been predicted to be observable in this regime, including modifications of both intensity, spectral features, and correlations of light-emitting devices with USC~\cite{Gambino2014, Genco2018}, as well as possible modifications of physical or chemical properties of systems ultrastrongly coupled to light~\cite{Ciuti2005, Ashhab2010, Galego2015, Herrera2016, Cirio2016, Kockum2017a}. This widespread interest led not only to the observation of the USC regime in a large number of physical implementations, but also to a steady increase of the normalized coupling, whose record is presently $\eta=1.43$~\cite{Bayer2017}.

The achievement of USC can be seen as the beginning of a third chapter in the history of light-matter interaction (see \figref{fig:SetupAndRegimes}). Already the control of this interaction afforded by the Purcell effect in the WC regime led to several important applications, e.g., low-threshold solid-state lasers~\cite{Vahala2003} and efficient single- and entangled-photon emitters~\cite{Shields2007, Salter2010}. Cavity QED with individual atoms in the SC regime made it possible to manipulate and control quantum systems, enabling both tests of fundamental physics~\cite{Haroche2013} and applications~\cite{Georgescu2012} such as high-precision measurements~\cite{Degen2017} and quantum information processing (QIP)~\cite{Wendin2017}. As the light-matter coupling strength reaches the USC regime, it starts to become possible to modify the very nature of the light and matter degrees of freedom. This opens new avenues for studying and engineering non-perturbatively coupled light-matter systems, which is likely to lead to novel applications. 

In this review, we gather both theoretical insights and experimental achievements in the field of USC. We begin by discussing various regimes of light-matter coupling in more detail, explaining their similarities and differences, the models used to describe them, and their properties. We then review how USC has been reached in different experimental systems. This is followed by an overview of defining characteristics of ultrastrong light-matter interaction such as virtual excitations and higher-order processes, topics which affect how the interaction of an USC system with an environment is treated. We also review quantum simulations of the USC regime, USC to a continuum instead of a single resonator mode, and how ultrastrong light-matter coupling is intimately connected to other areas of physics. We conclude with an outlook for the field, including possible new applications and outstanding challenges.


\section{Regimes and models for light-matter coupling}

The definitions of the WC, SC, and USC regimes compare the light-matter coupling strength $g$ to different parameters, as shown in \figref{fig:SetupAndRegimes}. Whether the coupling is strong or weak depends on whether $g$ is larger or not than the losses in the system. Ultrastrong coupling is not SC with larger couplings; its definition does not involve the value of losses but instead compares $g$ to bare energies in the system. It is thus possible for a system to be in the USC regime without having SC if losses are large~\cite{DeLiberato2017}. The ratio $\eta$ which defines USC instead determines whether perturbation theory can be used, and to what extent approximations can be made in models for the light-matter interaction.


\subsection{Models}

Some of the most fundamental models of light-matter interaction, the quantum Rabi, Dicke, and Hopfield models, are described in Box~1. However, these models, even though they do not approximate away some terms which are often ignored at low light-matter coupling strengths, still rely on some approximations, e.g., that the atoms are two-level systems and that the light is in a single mode. For ultrastrong light-matter coupling, these approximations may break down~\cite{DeBernardis2018, Stokes2018, SanchezMunoz2018}. 

As explained in Box~1, the light-matter interaction can be divided into two parts. It is essential to note that, in contrast to the terms in the first part (weighted by $g_1$), the terms in the second part (weighted by $g_2$) do \emph{not} conserve $\hat{N}_{\rm exc}$, the total number of excitations in the system. These latter terms are often referred to as anti-resonant or counter-rotating. When the light and matter frequencies are close to resonance, these terms can be omitted using the rotating-wave approximation (RWA). In the case of the quantum Rabi model (QRM), the RWA simplifies the Hamiltonian to the standard Jaynes--Cummings model (JCM)~\cite{Jaynes1963} (see Table B1.I). The JCM, which has been a workhorse of quantum optics in the WC and SC regimes, conserves $\hat{N}_{\rm exc} \equiv \hat a^\dagger \hat a + \sp \sm$ (symbols defined in Box~1) and restricts the resulting light-matter dynamics to two-dimensional Hilbert subspaces~\cite{Shore1993}. However, the RWA is not justified in the USC regime, when all terms in the light-matter interaction come into play.

\onecolumngrid

\begin{frshaded}

\section*{Box 1 --- Models for light-matter coupling}
\setcounter{equation}{0}
\renewcommand{\theequation}{B1.\arabic{equation}}

The quantum Rabi model~\cite{Rabi1937} (QRM) is a paradigm of quantum physics as one of the simplest and most fundamental models of light-matter interaction. In the QRM, the interaction between a single-mode bosonic field (e.g., a cavity mode with frequency $\omega_{\rm c}$) and a generic two-level system (TLS, or a qubit, with level splitting $\omega_{\rm q}$) is described by the quantum Rabi Hamiltonian ($\hbar=1$)
\bea
\hat H_{\rm Rabi} &=& \omega_{\rm c} \hat a^\dag \hat a + {\textstyle\frac{1}{2}} \omega_{\rm q} \sz + \hat H_{\rm int},
\label{eq:RabiModel}
\\
\hat H_{\rm int} &=& g \hat X \sx  =  g_1 \mleft( \hat a \sp + \hat a^\dag \sm \mright) +  g_2 \mleft( \hat a \sm + \hat a^\dag \sp \mright),
\label{eq:RabiInteraction}
\eea
where $\hat a$ ($\hat a^\dag$) is the annihilation (creation) operator of the cavity mode, $\sm = \ketbra{g}{e}$ ($\sp = \ketbra{e}{g}$) is the lowering (raising) operator between the ground ($\ket{g}$) and excited ($\ket{e}$) states of a given TLS, $\sx = \sm + \sp$ and $\sz = \ketbra{e}{e} - \ketbra{g}{g}$ are Pauli operators, and $\hat X = \hat a + \hat a^\dag$ is the canonical position operator of the electric field of the cavity mode. For simplicity, we ignore the vacuum-field energy in the free Hamiltonian in \eqref{eq:RabiModel}. Moreover, $g, g_1, g_2$ denote light-matter coupling strengths. In the QRM, $g = g_1 = g_2$, but this condition can be relaxed.

A Rabi-type model can also be applied to describe the interaction between two coupled harmonic oscillators. This is an effective description of many systems, where the light is coupled not to a single atom or molecule, but to an ensemble of these. For example, the standard fermion-boson QRM can be generalized to a purely bosonic multi-mode Hopfield model~\cite{Hopfield1958}, which describes the interaction between photons and collective excitations (e.g., plasmons or phonons) of a matter system. A simplified two-mode version of the Hopfield model is
\bea
\hat H_{\rm Hopfield} &=& \omega_{\rm c} \hat a^\dag \hat a + {\textstyle\frac{1}{2}} \omega_{\rm b} \hat b^\dag \hat b + \hat H'_{\rm int} + H_{\rm dia},
\label{eq:HopfieldModel}
\\
\hat H'_{\rm int} &=& g \hat X \hat Y' =  i g_1 \mleft( \hat a \hat b^\dag - \hat a^\dag \hat b \mright) + i g_2 \mleft(\hat a^\dag \hat b^\dag - \hat a \hat b \mright),
\label{eq:HopfieldInteraction}
\eea
where $\hat b$ ($\hat b^\dag$) is the annihilation (creation) operator for collective excitations of a matter system of frequency $\omega_{\rm b}$, and $\hat Y' = i (\hat b^\dag - \hat b)$ is the quadrature corresponding to the canonical momentum operator of the matter mode. The Hamiltonian $H_{\rm dia}$ describes the diamagnetic term (also referred to as the $A^2$ term), which is proportional to $\hat X^2$. This term is also sometimes added to the standard QRM. The physical meaning of $H_{\rm dia}$, and the conditions under which this term can be omitted, are explained in Box~2.

\begin{center}
\renewcommand{\arraystretch}{1.3}
\renewcommand{\tabcolsep}{0.15cm}
\begin{tabular}{ r | c | c }
 & 1 atom & $N$ atoms  \\
\hline 
no RWA & Quantum Rabi model~\cite{Rabi1937} & Dicke model~\cite{Dicke1954}, Hopfield model~\cite{Hopfield1958} \\
\hline
with RWA & Jaynes--Cummings model~\cite{Jaynes1963} & Tavis--Cummings model~\cite{Tavis1968} \\
\end{tabular}
\end{center}
{\small Table B1.I. The models used to describe various regimes of light-matter interaction.}

\end{frshaded}

\twocolumngrid


Although the QRM does not conserve $\hat{N}_{\rm exc}$, it does conserve the parity $\hat P = \exp (i \pi \hat{N}_{\rm exc})$. A generalized QRM, which is obtained by replacing the term $g \hat X \sx$ by $g \hat X (\sx \cos \theta + \sz \sin \theta)$ (with a parameter $\theta \neq 0, \pi$) does \emph{not} conserve even $\hat P$; this Hamiltonian features in experiments with superconducting circuits~\cite{Niemczyk2010, Yoshihara2017}. Note that the JCM conserves both $\hat{N}_{\rm exc}$ and $\hat P$.

An analytical approach to find the spectrum of the QRM was discovered only in 2011~\cite{Braak2011} (and has since been extended to multiple TLSs~\cite{Braak2013, Peng2013} and multiple bosonic modes~\cite{Chilingaryan2015}). But this solution is still based on conjectures and numerical calculations of transcendental (non-analytic) functions. A particular difficulty is to find exceptional eigenvalues of $\hat H_{\rm Rabi}$ with no definite parity (doubly degenerate)~\cite{Braak2011}. In contrast to the QRM, the spectrum of the JCM is simple and well-known~\cite{Shore1993}.

The QRM can be simulated with the standard JCM in experiments using various tricks, as discussed later in this review. Also, the coupling $g$ can be enhanced in various ways, e.g., by increasing the number of TLSs or cavity fields, or by applying classical (single-photon) drives to a single TLS or a cavity field. Recently, an exponential enhancement of the coupling $g$ was predicted with a two-photon drive (i.e., squeezing) of the cavity field~\cite{Qin2018, Leroux2018}.

A generalization of the QRM to $N$ TLSs (which can correspond to a single multi-level system or a large spin) is known as the Dicke model~\cite{Dicke1954}. Under the RWA, the Dicke model reduces to the Tavis--Cummings model~\cite{Tavis1968} (see Table B1.I). Another generalized version of the quantum Rabi model, with $g_1\neq g_2$, enables studying supersymmetry (SUSY), which exists if $g_1^2 - g_2^2 = \omega_{\rm c} \omega_{\rm q}$ (i.e., when the Bloch--Siegert shift~\cite{Bloch1940} is zero)~\cite{Tomka2015}. Note that $g_1 = g_2$ if the Rabi model is derived from first principles.

In Box~1, we give the Hamiltonian for the Hopfield model. In this case, the $g_1$ terms describe parametric frequency conversion, which conserves the total number of excitations $\hat{N}_{\rm exc}' \equiv \hat a^\dagger \hat a + \hat b^\dagger \hat b$, while the $g_2$ terms describe parametric amplification, which does \emph{not} conserve $\hat{N}_{\rm exc}'$. These processes, often studied in quantum optics, are analogous to those described by $H_{\rm int}$ for the QRM. This simplified Hopfield model has been applied to describe experimental data of a two-dimensional electron gas interacting with terahertz cavity photons in the USC regime~\cite{Hagenmuller2010, Zhang2016a, Li2018}.

\subsection{Other regimes of light-matter coupling}

For the sake of completeness, we here mention three other regimes of light-matter coupling which have been investigated in the literature. The first is the deep-strong-coupling [DSC, see \figref{fig:SetupAndRegimes}(f)] regime, in which $\eta$ becomes larger than one and higher-order perturbative processes are not only observable, but can become dominant. Theoretically investigated for the first time in 2010~\cite{Casanova2010}, this regime was finally demonstrated experimentally in 2017 using different physical implementations~\cite{Yoshihara2017, Bayer2017}.

The second, the very-strong-coupling (VSC) regime, is achieved when $g$ becomes comparable with the spacing between the excited levels of the quantum emitter. In this regime, although the number of excitations is conserved and first-order perturbation gives an adequate description of the system, the coupling is large enough to hybridize different excited states of the emitter, modifying its properties. This regime was initially predicted by Khurgin in 2001~\cite{Khurgin2001}, and observed in microcavity polaritons in 2017~\cite{Brodbeck2017}.

The third is the multi-mode-strong-coupling (MMSC), where $g$ exceeds the free spectral range of the resonator that the matter couples to. This regime has recently been reached with superconducting qubits coupled to either microwave photons in a long transmission-line resonator~\cite{Sundaresan2015} or phonons in a surface-acoustic-wave resonator~\cite{Moores2018}.

In the rest of this review, we will largely speak of USC, with the implicit understanding that, according to the value of $\eta$ and other energy scales, the system under investigation could also be in the WC, SC, VSC, MMSC, or DSC regimes.


\section{Properties of ultrastrongly coupled systems}

As $\eta$ increases, several properties of coupled light-matter systems change drastically. In \figref{fig:EnergyLevelsGroundStateUSC}(a), we plot, as a function of $\eta$, the lowest energy levels of a light-matter system with a single atom on resonance with a cavity mode. Only the quantum Rabi model (see Box~1) gives a correct picture of the energy levels for all $\eta$; various approximate methods can be used for small or large $\eta$. The Jaynes--Cummings model correctly predicts the Rabi splitting (dressed states) between neighboring pairs of energy levels, but fails when the system enters the USC regime.


\subsection{Ground-state properties}

\begin{figure*}
\centering
\includegraphics[width=\linewidth]{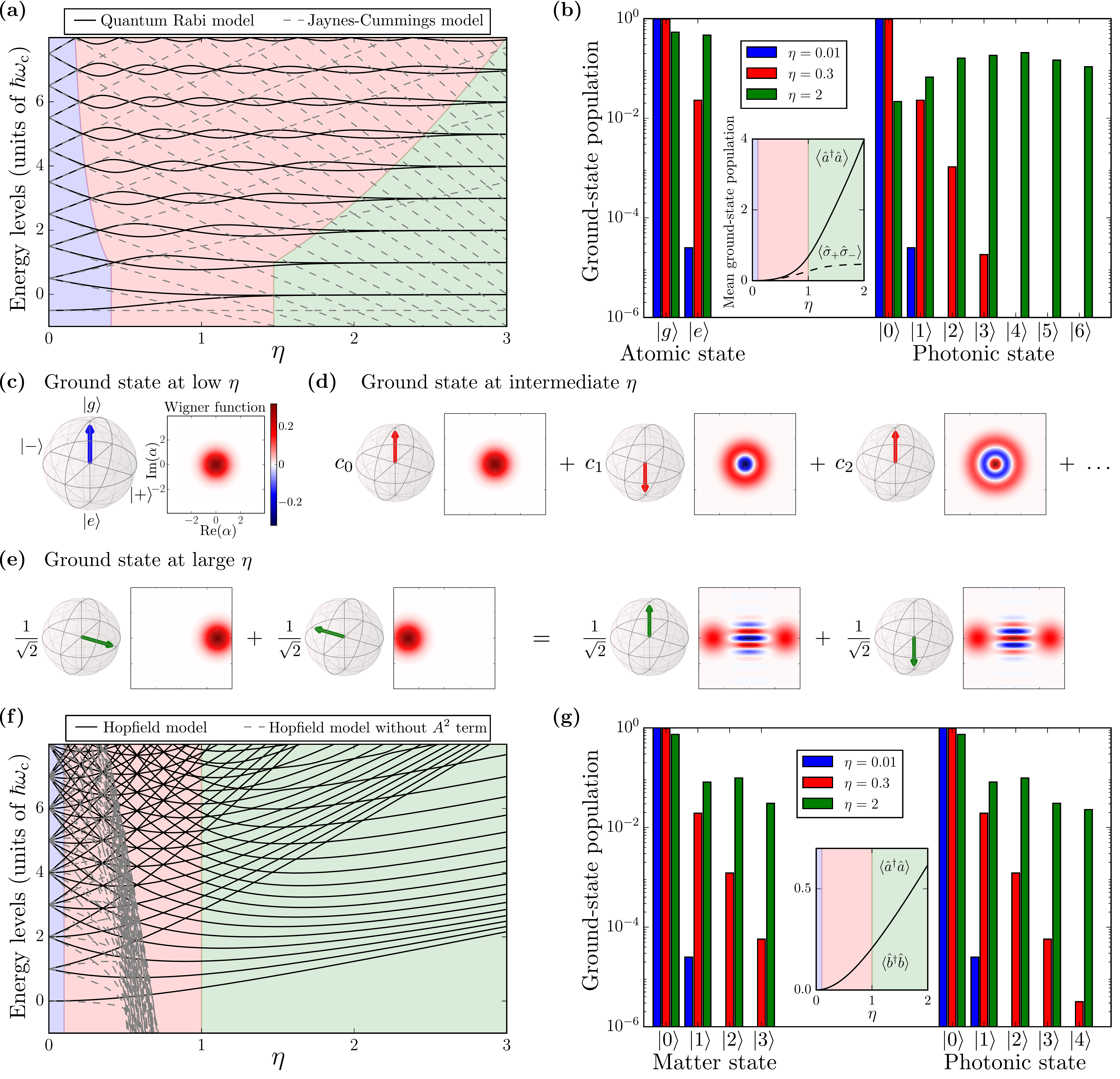}
\caption{Spectrum and ground-state properties of ultrastrongly coupled light-matter systems.
\textbf{(a)} The lowest energy levels (offset by $+\hbar g^2/\omega_{\rm c}$) of the quantum Rabi Hamiltonian (solid black curves) as a function of the normalized coupling strength $\eta$. For comparison, we also plot the same levels for the Jaynes-Cummings Hamiltonian (dashed grey curves). The two models coincide in the shaded blue area, where perturbation theory using $\eta$ as a small parameter works well. The cross-over into the shaded red area, where perturbation theory no longer works, takes place around the Juddian points, where pairs of energy levels begin to cross. Once $\eta$ is well past 1 (shaded green area), other parameters become comparatively small and can be used for perturbative expansions. The cross-over to this regime is marked by pairs of energy levels starting to become degenerate~\cite{Ashhab2010, Nataf2010a}. More details can be found in Ref.~\cite{Rossatto2017}, which inspired this figure.
\textbf{(b)} Representative ground-state statistics for the SC (blue), USC (red), and DSC (green) regimes of the QRM. As the coupling increases, the ground state of the QRM starts to contain a significant number of (virtual) atomic and photonic excitations. In the inset, the shaded blue, red, and green areas indicate non-USC, USC, and DSC, respectively, in the usual convention where USC encompasses $0.1 < \eta <1$.
\textbf{(c)} The ground state of the QRM for low coupling strength is well approximated as $\ket{E_0} \approx \ket{g}\ket{0}$, here illustrated with a Bloch sphere representation for the atomic state and a Wigner-function representation for the photonic state.
\textbf{(d)} In the non-perturbative USC regime, no simple expression for the ground state of the QRM exists. It is a superposition of all states with an even number of excitations: $\ket{E_0} = c_0 \ket{g}\ket{0} + c_1\ket{e}\ket{1} + c_2\ket{g}\ket{2} + c_3\ket{e}\ket{3} + \ldots$, where, as can be seen in (b), $\abssq{c_0} > \abssq{c_1} > \ldots$.
\textbf{(e)} As the coupling is increased further, into the perturbative DSC regime, the QRM ground state can be approximated well as $\ket{E_0} \approx \left( \ket{+}\ket{\alpha} + \ket{-}\ket{-\alpha} \right) / \sqrt{2}$, where $\ket{\pm} = \left( \ket{g} \pm \ket{e} \right) / \sqrt{2}$ are the eigenstates of $\sx$ and $\ket{\alpha}$ is the coherent state with $\alpha = \sqrt{\expec{\hat a^\dag \hat a}}$. An interesting feature of this ground state is that it can be rewritten as $\left( \ket{g} \left[ \ket{\alpha} - \ket{-\alpha} \right] / \sqrt{2} + \ket{e} \left[ \ket{\alpha} + \ket{-\alpha} \right] / \sqrt{2} \right) / \sqrt{2}$, i.e., the atom is entangled with photonic Schr\"{o}dinger-cat states.
\textbf{(f)} Same as (a), but for the Hopfield model with (solid black curves) and without (dashed grey curves) the $A^2$ term. For the latter case, the sharp drop-off beginning at $\eta = 0.5$ marks the superradiant phase transition.
\textbf{(g)} Same as (b), but for the Hopfield model. 
\label{fig:EnergyLevelsGroundStateUSC}}
\end{figure*}

The difference between the USC and non-USC regimes is particularly striking for the ground state of the coupled light-matter system, as shown in \figref{fig:EnergyLevelsGroundStateUSC}(b)-(e). For small $\eta$, the lowest-energy state of the system is simply an empty cavity with the atom in its ground state. However, as $\eta$ grows the coupling makes it increasingly energetically favorable to have atomic and photonic excitations in the ground state. The exact nature of these excitations is discussed later in this review, in the section on virtual excitations. Here we only note that for very large $\eta$, in the DSC regime, as shown in \figref{fig:EnergyLevelsGroundStateUSC}(e), the ground state of the QRM consists of photonic Schr\"{o}dinger-cat states entangled with the atom and exhibits nonclassical properties such as squeezing~\cite{Hines2004, Ashhab2010}.

As shown in the inset of \figref{fig:EnergyLevelsGroundStateUSC}(b), the mean number of photons in the ground state starts to increase rapidly when $\eta$ approaches and passes one. In the case of many atoms coupled to the light, as described by the Dicke model (see Box~1), it is predicted that a quantum phase transition, known as the superradiant phase transition~\cite{Hepp1973, Wang1973, Emary2003} takes place at a critical value of $\eta$, separating phases with and without photons in the ground state of the system. However, as explained below, whether or not this phase transition actually occurs depends on whether an additional term, the diamagnetic term, should be included in the Hamiltonian.

In \figref{fig:EnergyLevelsGroundStateUSC}(f), we plot the energy levels for the case where the matter instead consists of many atoms and is described as bosonic collective excitations in the Hopfield model (see Box~1). The impact of the diamagnetic term is clearly seen here. In \figref{fig:EnergyLevelsGroundStateUSC}(g), we plot the ground-state population in the same way as in \figref{fig:EnergyLevelsGroundStateUSC}(b) with the diamagnetic term included. Also in this case, the ground state contains virtual light and matter excitations. This ground state can be calculated analytically~\cite{Quattropani1986} for all $\eta$; it is a multi-mode squeezed state for large $\eta$.


\subsection{The diamagnetic term}

In Box~2, we provide more details on the diamagnetic term. In the DSC regime, the diamagnetic term can act as a potential barrier for the photonic field, localizing it away from the dipoles, leading to an effective decoupling between the light and matter degrees of freedom~\cite{DeLiberato2014, Garcia-Ripoll2015}. This means that the Purcell effect, known from the WC regime and thought to increase the spontaneous-emission rate of the qubit as $g$ increases, actually becomes negligible when $g$ becomes large enough~\cite{DeLiberato2014}. A similar decoupling can occur if qubit-qubit interactions are added to the Dicke model~\cite{Jaako2016}. Even in the pure QRM, unexpected changes in photon-output statistics take place deep in the DSC regime~\cite{LeBoite2016}.

\onecolumngrid

\begin{frshaded}

\section*{Box 2 --- The diamagnetic term}
\setcounter{equation}{0}
\renewcommand{\theequation}{B2.\arabic{equation}}

The minimal-coupling substitution $\hat{p} \rightarrow \hat{p} - e \hat{A}$ (where $\hat{p}$ is the momentum, $e$ the elementary charge, and $\hat{A}$ the electromagnetic vector potential) in the kinetic Hamiltonian $\hat{H}_{\text{kin}} = \frac{\hat{p}^2}{2m}$ (where $m$ is the mass of the charged particle) leads, when expanding the square, to the appearance of two interaction terms. The first,
\be
\hat{H}_{\text{int}} = -\frac{e \hat{p} \hat{A}}{m} = \sum_n g_{jn} \mleft(\hat{a} + \hat{a}^\dag \mright) \hat{M}_{jn},
\ee
is of the form considered in Box~1, describing a dipolar interaction between the photonic cavity mode and the optically active transitions between the initial state $j$ and all final states $n$, with $\hat{M}_{jn}$ a generic transition operator. The second term,
\be
\hat{H}_{\text{dia}} = \frac{e^2 \hat{A}^2}{2m} = D \mleft( \hat{a} + \hat{a}^\dag \mright)^2,
\ee
is the one responsible for the appearance of diamagnetism and, being of second order in the electric charge, is usually of limited importance when studying dipolar transitions outside the USC regime. 

The link between the intensity of the diamagnetic term $D$, which does not depend on the dipolar matrix element, and the strength of the light-matter coupling can be highlighted by exploiting the Thomas--Reiche--Kuhn sum rule, allowing to rewrite $\hat{H}_{\text{dia}}$ as
\be
\hat{H}_{\text{dia}} = \sum_n \frac{\abssq{\brakket{n}{\hat{p}}{j}}}{\hbar\omega_{jn}}
\frac{e^2 A(\mathbf{r})^2}{m^2} = \sum_n \frac{\hbar g_{jn}^2}{\omega_{jn}} \mleft(\hat{a} + \hat{a}^\dag \mright)^2,
\label{eq:Hdia}
\ee
with $\omega_{jn}$ the frequency of the $j \rightarrow n$ transition. From \eqref{eq:Hdia} it is clear that when any single quasi-resonant transition of frequency $\omega_x$ and coupling $g$ is considered,
\be
D\geq \frac{g^2}{\omega_x}=g\eta.
\label{eq:Dvalue}
\ee
The ratio of the coefficients of the diamagnetic and dipolar parts of the light-matter interaction, $D/g$, is thus at least as large as the normalized coupling $\eta$, with the equality in \eqref{eq:Dvalue} if a single transition saturates the sum rule. The impact of the diamagnetic term is thus non-negligible in the USC regime, and eventually becomes dominant in the DSC regime. 

\end{frshaded}

\twocolumngrid


Being a consequence of gauge invariance, the diamagnetic term is required to obtain a consistent theory, including in superconducting systems. Claims have been advanced on the possibility to engineer systems, both dielectrics~\cite{Hagenmuller2012} and superconducting~\cite{Nataf2010}, in which the diamagnetic term is absent or at least reduced to violate~\eqref{eq:Dvalue}. Those claims have attracted strong criticism~\cite{Viehmann2011, Chirolli2012} and there is for the moment no consensus on this point. There have also being theoretical proposals showing how the matter could be experimentally settled~\cite{Tufarelli2015, Rossi2017}.

The historically most important role played by the diamagnetic term in CQED is linked with a series of no-go theorems~\cite{Rzazewski1975, Slyusarev1979, Viehmann2011, Bamba2014} seemingly demonstrating that its presence makes a system stable against superradiant phase transitions. Although we will not dwell on the details of what is a very subtle, and still debated, topic, in which attention has to be paid to the specificities of each model~\cite{Keeling2007, Vukics2012, Baksic2013, Vukics2014, Bamba2017}, it is easy to gain an intuitive understanding of why such a term has been predicted to stabilize the system ground state. The diamagnetic term can in fact be removed from the Hamiltonian by performing a Bogoliubov rotation in the space of the photon operator, at the cost of a renormalization of the cavity frequency: $\omega_{\rm c} \rightarrow \sqrt{\omega_{\rm c}^2 + 4 \omega_{\rm c} D}$. In order for the system to undergo a quantum phase transition, the coupling has to be strong enough to push one of the system eigenmodes to zero frequency. The blue-shift of the cavity due to the renormalization thus implies a larger coupling $g$ is required to reach the critical point, but from \eqref{eq:Dvalue} this in turn will further blue-shift the cavity mode. A careful calculation shows that, at least for the Dicke model described in Box~1, this runaway process leads to a divergent critical value of $g$ if \eqref{eq:Dvalue} holds.

Although the diamagnetic part of the Hamiltonian is often referred to as $A^2$ term, it can take different forms under a unitary transformation. Using, e.g., the Power-Zienau-Woolley form of the Hamiltonian, the term $A^2$ disappears, being substituted by an equivalent $P^2$ term~\cite{Todorov2012, Vukics2012}, quadratic in the matter instead of in the photonic field. Finally, it is worth noting that the presence of a squared field term in the Hamiltonian, assuring the stability of matter linearly coupled to a bosonic field, is a feature not limited to the interaction with the transverse electromagnetic field. Similar terms, satisfying the equivalent of \eqref{eq:Dvalue}, have, in particular, been derived in the case of longitudinal interactions in intersubband polarons~\cite{DeLiberato2012}.


\section{Experimental systems with ultrastrong coupling}

\begin{figure*}
\centering
\includegraphics[width=\linewidth]{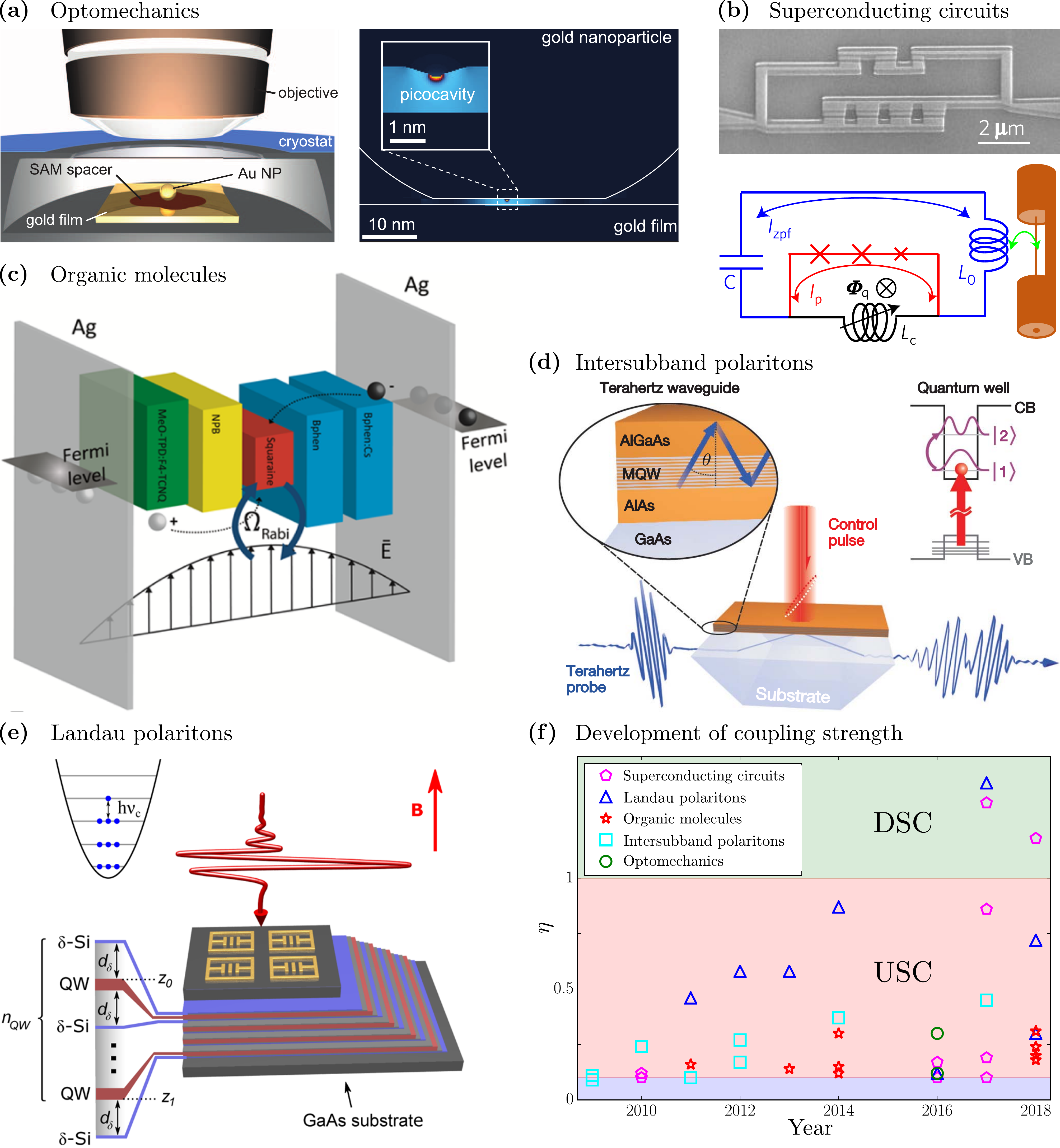}
\caption{Experimental systems with ultrastrong light-matter coupling.
\textbf{(a)} Optomechanics. In Ref.~\cite{Benz2016}, vibrational modes of biphenyl-4-thiol molecules in a self-assembled monolayer (SAM) interacted with light localized to a ``picocavity'' formed between a gold film and a single gold atom on the surface of a gold nanoparticle, reaching $\eta = 0.3$.
\textbf{(b)} Superconducting circuits. In Ref.~\cite{Yoshihara2017}, a flux qubit consisting of three Josephson junctions in a loop (image in upper panel; red part of the sketch in the lower panel) coupled inductively to a lumped-element $LC$ circuit, reaching $\eta = 1.34$.
\textbf{(c)} Organic molecules. In Ref.~\cite{Gambino2014}, squaraine dye was placed in-between layers of organic materials (p- and n-doped layers and optical spacers) in a microcavity formed by silver mirrors, forming an organic light-emitting diode reaching $\eta = 0.3$.
\textbf{(d)} Intersubband polaritons. In Ref.~\cite{Gunter2009}, a transition between subbands $\ket{1}$ and $\ket{2}$ in the conductance band (CB) of multiple quantum wells (MQWs) was activated by a near-infrared control pulse exciting electrons from the valence band (VB). The intersubband transition coupled to TM-polarized cavity photons propagating at an angle $\theta$, resulting in $\eta = 0.09$.
\textbf{(e)} Landau polaritons. In Ref.~\cite{Bayer2017}, a stack of quantum wells, hosting 2DEGs with Landau levels (set by an external magnetic field $\mathbf{B}$) separated by the cyclotron frequency $\nu_{\rm c}$, were coupled to an array of THz resonators on top of the stack, reaching $\eta = 1.43$.
\textbf{(f)} Measured $\eta$ for all experiments that have achieved USC, excluding experiments with USC to a continuum and quantum simulations of USC. In something that can be called ``Moore's law for light-matter coupling strength'', the past decade has seen experiments progressing steadily from breaking the barrier to the USC regime to entering the DSC regime.
Figures reproduced with permission from: \textbf{(a)} Ref.~\cite{Benz2016} \copyright 2016, AAAS; \textbf{(b)} Ref.~\cite{Yoshihara2017} \copyright 2017, NPG; \textbf{(c)} Ref.~\cite{Gambino2014}, \copyright 2014, ACS; \textbf{(d)} Ref.~\cite{Gunter2009} \copyright 2009, NPG; and \textbf{(e)} Ref.~\cite{Bayer2017} \copyright 2017, ACS.
\label{fig:ExperimentsUSC}}
\end{figure*}

The first experimental demonstration of ultrastrong ($\eta > 0.1$) light-matter coupling was reported in 2009~\cite{Anappara2009}. As shown in \figref{fig:ExperimentsUSC} and explained below, USC has since been achieved in several different systems and at different wavelengths of light. In 2017, two experiments even managed to reach DSC ($\eta > 1$)~\cite{Yoshihara2017, Bayer2017}. The past decade has seen a rapid increase of $\eta$ [\figref{fig:ExperimentsUSC}(f)].

\subsection{Intersubband polaritons}

The USC regime was first predicted~\cite{Ciuti2005} and demonstrated~\cite{Anappara2009} exploiting intersubband polaritons in microcavity-embedded doped quantum wells. In these systems, nanoscopic layers of different semiconductors create a confining potential for carriers along the growth direction, which splits electronic bands into discrete parallel subbands. Thanks to the quasi-parallel in-plane dispersion of the different conduction subbands, all the electrons in the conduction band can be coherently excited, creating narrow collective optical resonances. The coupling of these resonances with transverse-magnetic (TM)-polarized radiation scales with the square root of the total electron density. By modifying the width of the quantum wells, the resonances can be tuned to cover the THz and mid-infrared sections of the electromagnetic spectrum. 

Intersubband-polariton systems are usually well described by the Dicke model (see Box~1). This was exploited in Ref.~\cite{Anappara2009}, where demonstration of USC with $\eta = 0.11$ was obtained by comparing experimental data with best fittings obtained using the Dicke model with and without anti-resonant terms. 

However, this appealing simple model can be spoiled in more complex devices. The presence of multiple quasi-resonant photonic modes can in fact lead instead to a physics described by the quantum Rabi model~\cite{DeLiberato2013} (see Box~1). Moreover, as the quantum-well width increases and multiple electron transitions become available, the intuitive picture in terms of single-particle states is lost. In that case, the electronic transition is better described as a plasma-like mode named after Berreman~\cite{Askenazi2014, Askenazi2017}.

Intersubband polaritons remain a scientifically and technologically interesting system to study USC phenomenology thanks to the possibility to non-adiabatically modify the coupling strength~\cite{Gunter2009}, making it a promising platform for quantum vacuum-emission experiments~\cite{DeLiberato2007, Auer2012}. Moreover, $\eta$ has been progressively increased in various experiments~\cite{Todorov2010, Jouy2011, Geiser2012, Delteil2012, Askenazi2014} up to the present value $\eta = 0.45$~\cite{Askenazi2017}.


\subsection{Superconducting circuits}

The next experiments to reach USC, in 2010~\cite{Niemczyk2010, Forn-Diaz2010}, used superconducting quantum circuits (SQCs). In these systems, electrical circuits with Josephson junctions, operating at GHz frequencies, function as ``artificial atoms'', acquiring a level structure similar to that of natural atoms when cooled to millikelvin temperatures. These artificial atoms are then coupled to photons in $LC$ or transmission-line resonators. Superconducting circuits are a powerful platform for exploring atomic physics and quantum optics, and for QIP, since their properties (resonance frequencies, coupling strength, etc.) can be designed and even tuned in situ~\cite{Gu2017}. This has already been widely exploited in the SC regime to, e.g., engineer quantum states and realize quantum gates.

The SQC experiments~\cite{Niemczyk2010, Forn-Diaz2010, Baust2016, Forn-Diaz2016, Bosman2017, Chen2017, Yoshihara2017, Yoshihara2017a, Yoshihara2018} are the only ones that have achieved USC with a \textit{single} (albeit artificial) atom. The reason that superconducting circuits, unlike other experimental systems, do not require collective excitations to reach USC, is that the coupling scales differently with $\alpha$ in these circuits~\cite{Devoret2007}. In cavity QED, the coupling scales as $\alpha^{3/2}$. However, in circuit QED, the coupling scales as either $\alpha^{1/2}$ or $\alpha^{-1/2}$, depending on the layout of the superconducting circuit.

The design used in Ref.~\cite{Yoshihara2017}, which by reaching $\eta = 1.34$ was the first to break the DSC barrier, is shown in \figref{fig:ExperimentsUSC}(b). As discussed in more detail later in this review, SQCs are also the only systems where USC to a continuum~\cite{Forn-Diaz2017, Magazzu2018, PuertasMartinez2018} and quantum simulation of USC~\cite{Langford2017, Braumuller2017} has been demonstrated.


\subsection{Landau polaritons}

Since 2011, the record for $\eta$ has almost continuously been held by Landau-polariton systems. In these systems, based on microcavity-embedded doped quantum wells under a transverse magnetic field, the USC occurs between a photonic resonator and the collective electronic transitions between continuous Landau levels. Contrary to intersubband polaritons, whose dipole lies along the growth axis, Landau transitions have an in-plane dipole and thus couple to transverse-electric-polarized radiation. The very large coupling achievable in these systems is due to an interplay between the degeneracy of Landau levels, the transition dipole which increases with the index of the highest occupied Landau level, and the relatively small cyclotron frequencies in the THz or GHz range observable in high-mobility heterostructures.

Theoretically described for the first time in 2010~\cite{Hagenmuller2010}, Landau-polariton systems with USC were observed shortly afterward using split-ring resonators~\cite{Scalari2012, Scalari2013, Maissen2014, Bayer2017, Keller2017} [see \figref{fig:ExperimentsUSC}(e)], photonic-crystal cavities~\cite{Zhang2016a, Li2018}, and coplanar microresonators~\cite{Muravev2011}. The present world-record value of light-matter coupling, $\eta=1.43$, was measured in Ref.~\cite{Bayer2017}.

Landau-polariton systems have revealed themselves to be a useful platform for investigating USC phenomenology. In Ref.~\cite{Li2018}, the polarization selectivity of the Landau transition was used to directly measure the Bloch-Siegert shift due to the antiresonant terms in the Hamiltonian. Furthermore, in Ref.~\cite{Paravicini-Bagliani2018}, magnetotransport was used to investigate the nature of the matter excitations participating in the Landau-polariton formation. In Ref.~\cite{Bayer2017}, light-matter decoupling in the DSC regime was observed for the first time, and also exploited to optimize the design of the photonic resonator.


\subsection{Organic molecules}

The USC regime has also been realized at room temperature at a variety of optical frequencies, coupling cavity photons (or, in one case, plasmons~\cite{Todisco2018}) to Frenkel molecular excitons \cite{Schwartz2011, Kena-Cohen2013, Gambino2014, Gubbin2014, Mazzeo2014, Genco2018, Barachati2018, Eizner2018}. These systems consist of thin films of organic molecules with giant dipole moments (which make it possible to reach USC) sandwiched between metal mirrors [see \figref{fig:ExperimentsUSC}(c)] and present an interesting combination of high coupling strengths and functional capacities. A vacuum Rabi splitting beyond \unit[1]{eV}, corresponding to $\eta = 0.3$, has been reported~\cite{Gambino2014, Barachati2018}. Using such high coupling strengths, monolithic organic light emitting diodes (OLED) working in the USC regime have been made~\cite{Gubbin2014, Mazzeo2014, Gambino2014, Genco2018, Eizner2018}. These devices exhibit a room-temperature dispersion-less angle-resolved electroluminescence with very narrow emission lines that can be exploited to realize innovative optoelectronic devices.


\subsection{Optomechanics}

The concept of ultrastrong light-matter interaction can be extended to optomechanics. Recently, the USC limit was reached in a setup where plasmonic picocavities interacted with the vibrational degrees of freedom of individual molecules~\cite{Benz2016} [see \figref{fig:ExperimentsUSC}(a)], achieving $\eta = g/ \omega_{\rm m} = 0.3$ ($\omega_{\rm m}$ is the mechanical frequency). The increase in coupling strength here is due to the small mode volume of the picocavity, which circumvents the diffraction limit to confine optical light in a volume measured in cubic nanometers.

Another approach to increase optomechanical coupling strength is to use molecules with high vibrational dipolar strength (similar to the preceding subsection). This was the approach in Ref.~\cite{George2016a}, which reached $\eta = 0.12$. The USC limit has also been approached in circuit-optomechanical systems by using the nonlinearity of a Josephson-junction qubit to boost $\eta$~\cite{Pirkkalainen2015}.


\section{Virtual excitations}

As shown above in \figref{fig:EnergyLevelsGroundStateUSC}, a clear difference between USC systems and those with lower coupling strength is the presence of light and matter excitations in the ground state. This difference is due to the influence of the counter-rotating terms in the system Hamiltonian (Box~1). At lower coupling strength, excited states of the system can be ``dressed states'', superpositions of two states containing both light and matter excitations~\cite{Shore1993}. These two states contain the \textit{same} number of excitations. However, in the USC regime, all excited states are dressed by \textit{multiple} states containing \textit{different} numbers of excitations. Much research on USC systems has dealt with understanding whether these excitations dressing the system states (especially the ground state) are real or virtual, how these excitations can be probed or extracted, how they make possible higher-order processes that mirror nonlinear optics~\cite{Kockum2017a} (see also the section on applications below), and how they affect the description of input and output for the system (Box~3).


\subsection{Dressed states and input-output theory}

As explained in more detail below and in Box~3, a correct treatment of input-output, decoherence, and correlation functions for a USC system, requires taking into account that the system operators coupling the system to the outside world no longer induce transitions between the bare states of the system (which have fixed numbers of photons and atomic excitations). Instead, the transitions are between the dressed, true eigenstates of the system (which contain contributions from various numbers of photons and atomic excitations)~\cite{Ashhab2010, Beaudoin2011, Ridolfo2012, Stassi2016}. Following the development of such a treatment, several interesting properties of USC systems have been revealed. For example, while thermal emission of photons is supposed to be bunched (photons tend to be emitted together) and photon emission from a single atom is supposed to be anti-bunched (photons are emitted one by one), the photons emitted from a thermalized cavity in the USC regime can be anti-bunched~\cite{Ridolfo2013} and a two-level atom coupled ultrastrongly to a cavity can emit bunched photons~\cite{Garziano2017}.


\onecolumngrid

\begin{frshaded}

\section*{Box 3 --- Treating open quantum systems in the USC regime}
\setcounter{equation}{0}
\renewcommand{\theequation}{B4.\arabic{equation}}

No quantum system is completely isolated from its environment. Control and readout imply a coupling with the outside world, leading to dissipation and decoherence. Textbook quantum-optical procedures to treat {\em open} quantum systems neglect the interaction between their constituent subsystems when describing their coupling to the environment~\cite{Gardiner2004} (see Fig.~B3). This results in a set of dynamical equations where each subsystem couples to the environment at its own bare frequency. This {\it white-reservoir} approximation fails dramatically in the USC regime, as it does not take into account that the environment density of states  vanishes at negative frequencies. The large frequency shifts of the USC regime in fact push part of the system's spectral weight to negative frequencies. A frequency-independent environment density of states thus allows for coupling the system with negative-frequency modes, making the ground state unstable even at zero temperature.

This problem was solved in Ref.~\cite{DeLiberato2009}, by using a master equation with colored reservoirs. A general approach to project the master equation on the coupled eigenbasis was then developed in Ref.~\cite{Beaudoin2011} and it has been the subject of various other works~\cite{Ashhab2010, Ridolfo2012, Bamba2012, Bamba2014a, Stassi2016, Bamba2016}.
Numerical simulation of the resulting master equations can become computationally demanding for larger values of $\eta$, because the increasing number of virtual photons requires exponentially larger simulation cutoffs. 

A method both analytically simpler and numerically lighter is the bosonic input-output theory, in which the system dynamics is integrated out to derive the scattering matrix. This approach was initially introduced for the USC regime in Ref.~\cite{Ciuti2006}, and further developed in following works~\cite{Bamba2012, Bamba2013, DeLiberato2014a}. Although applicable only to bosonic quadratic Hamiltonians, this approach has the advantage of being non-perturbative. It thus enables investigating loss-dominated regimes, allowing, e.g., to study the impact of the environment on the population of ground-state virtual excitations~\cite{DeLiberato2017}.

\begin{center}
\includegraphics[width=0.8\linewidth]{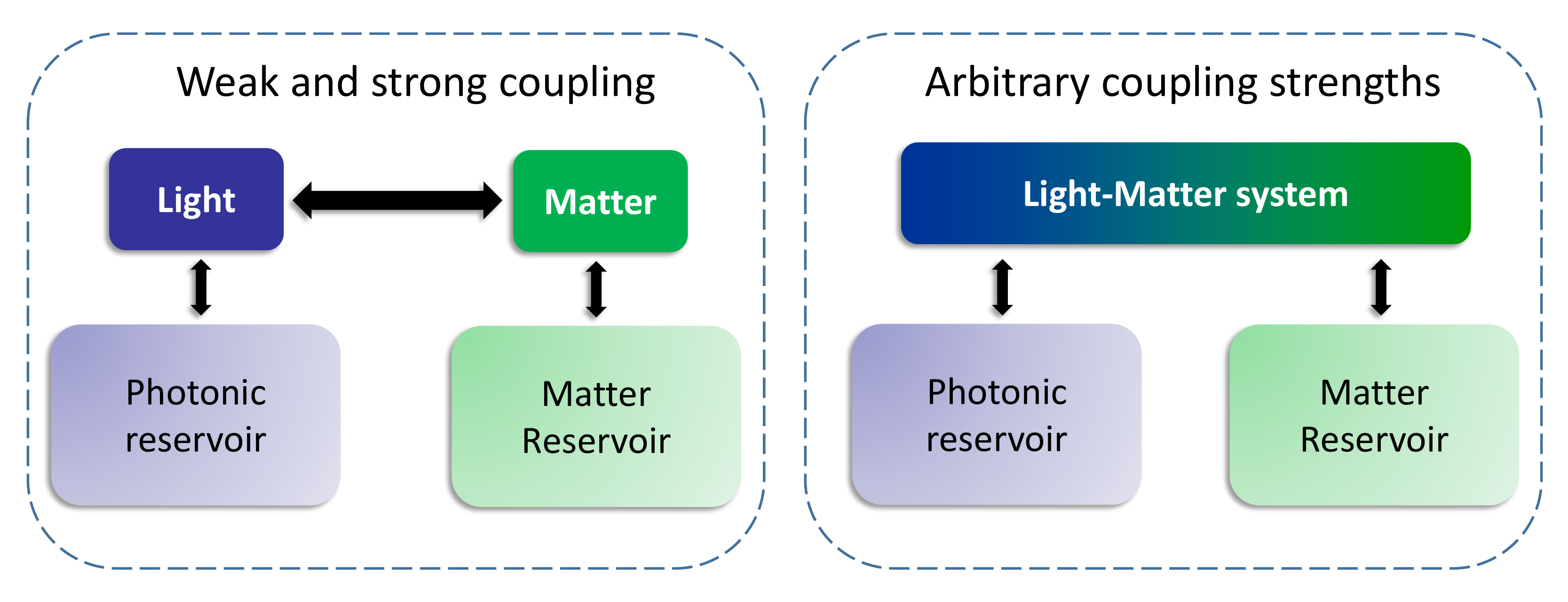}
\end{center}
{\small Figure B3. An illustration of the difference between open quantum systems without (left) and within (right) the USC regime. As the light-matter coupling strength increases, it becomes necessary to describe the interaction with the environment in terms of the coupled eigenmodes of the light-matter system.}

\end{frshaded}

\twocolumngrid


A simple way to understand the issue of open quantum systems in the USC regime is to remember that since the Hamiltonian of such a system is non-number conserving, its ground state contains a finite population of virtual excitations (see \figref{fig:EnergyLevelsGroundStateUSC}). Assuming that the emitted radiation is just proportional to the photon population in the cavity, neglecting to discriminate between real and virtual particles, leads to the prediction of unphysical radiation from its ground state~\cite{Ciuti2006, Ridolfo2012}. As first shown for confined polaritons~\cite{Savasta1996}, the quantum operators that correctly describe the emission of an output photon in the USC regime contain contributions from both \textit{bare} annihilation and \textit{bare} creation cavity-photon operators.

As shown in Ref.~\cite{Ridolfo2012}, the resulting input-output relation contains the positive-frequency operator $\hat X^+ = \sum_{i < j} X_{ij} \ketbra{E_i}{E_j}$ instead of the cavity-mode annihilation operator $a$. Here, $\ket{E_i}$ are the dressed eigenstates of the ultrastrongly coupled system, ordered such that $E_j  > E_i$ for $j > i$. The coefficients $X_{ij}$ are matrix elements between eigenstates. In the simplest case: $X_{ij} = \brakket{E_i}{\hat a + \hat a^\dag}{E_j}$. The operator $\hat X^+$ can be interpreted as the operator describing the annihilation of {\em physical} photons in the interacting system. Analogously, $\hat X^- \equiv (\hat X^+)^\dag$ corresponds to the creation operator. It is interesting to note that, while in the ground state $\ket{E_0}$ of a system in the USC regime the number of {\em bare} photons is nonzero, $\brakket{E_0}{a^\dag a}{E_0} \neq 0$ (see \figref{fig:EnergyLevelsGroundStateUSC}), the definition of $\hat X^+$ automatically implies that the number of detectable photons is zero: $\brakket{E_0}{\hat X^- \hat X^+}{E_0} = 0$.


\subsection{Probing and extracting virtual photons}

\begin{figure*}
\centering
\includegraphics[width=\linewidth]{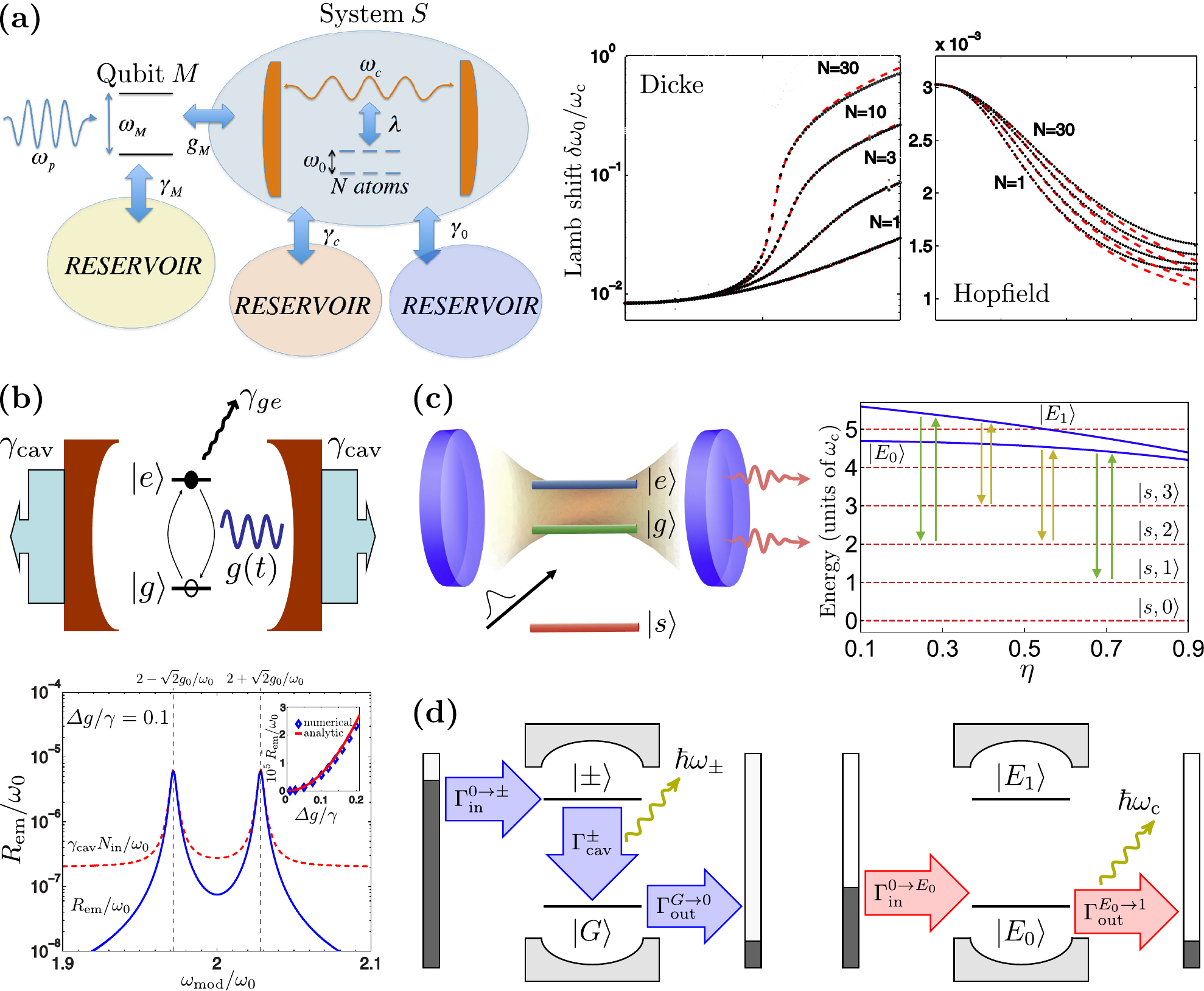}
\caption{Proposed methods for probing and extracting virtual photons dressing the states of a USC system. 
\textbf{(a)} To probe the ground state of a system $S$ ($N$ atoms with frequencies $\omega_0$, coupled with strength $\lambda$ to a cavity mode with frequency $\omega_{\rm c}$) described by the quantum Rabi, Dicke, or Hopfield models (see Box~1), Ref.~\cite{Lolli2015} proposes to connect an ancillary qubit $M$ (transition frequency $\omega_{\rm M}$) to the cavity, as sketched to the left (the $\gamma$ are relaxation rates). The coupling $g_{\rm M}$ is not ultrastrong. As shown to the right for the Dicke and Hopfield models, the Lamb shift of qubit $M$ depends on $N$ and $\eta$ for the system. One way to read out this shift is to send in a probe tone at a frequency $\omega_{\rm p}$ and monitor the population of qubit $M$. Note that qubit M cannot absorb any ground-state photons from the system, since they are bound there~\cite{DiStefano2017a}.
\textbf{(b)} Virtual photons in the ground state of a USC system can be released by modulating the coupling $g(t)$ around its original value $g_0$ at a frequency $\omega_{\rm mod}$ with an amplitude $\Delta g$, as schematically shown in the upper panel~\cite{DeLiberato2009}. As shown in the lower panel, the photon emission rate $R_{\rm em}$ (blue solid curve) has two peaks close to $\omega_{\rm mod} = 2 \omega_0$ ($\omega_0$ denotes the cavity and qubit frequencies).  The inset shows the emission rate at one of the peaks as a function of $\Delta g$ (scaled by the relaxation rates $\gamma$). Note that this calculation requires the proper treatment of input-output theory for USC discussed in Box~3. The red dashed curve shows the result if one uses the standard theory; this predicts unphysical photon emission at all modulation frequencies, proportional to $N_{\rm in}$, the number of intracavity photons.
\textbf{(c)} Virtual photons can also be extracted from a USC system through stimulated transitions. The left panel depicts a situation where a three-level atom has its upper transition ($\ket{g} \leftrightarrow \ket{e}$) coupled ultrastrongly to a cavity mode. The blue solid curves in the right panel show the first and second energy levels for the ultrastrongly coupled part of this system. However, since the transitions $\ket{s} \leftrightarrow \ket{g}$ and $\ket{s} \leftrightarrow \ket{e}$ are not ultrastrongly coupled to the cavity, there may be states $\ket{s,n}$ (atom in $\ket{s}$ and $n$ photons in the cavity) that have lower energy. Stimulating (black arrow in the left panel) a transition $\ket{s} \leftrightarrow \ket{g}$ (green arrows) or $\ket{s} \leftrightarrow \ket{e}$ (yellow arrows) can thus release or absorb $n$ photons in the cavity~\cite{DiStefano2017}.
\textbf{(d)} Yet another way to stimulate the release of photons from a USC ground state is electroluminescence. In standard electroluminescence, depicted in the left panel, a current flowing at a rate $\Gamma$ from the reservoir on the left through an electronic two-level system coupled to a photonic cavity can release a photon if the electron occupies an excited state $\ket{\pm}$ and then relaxes to the ground state $\ket{G}$ before passing to the reservoir on the right. However, as shown in the right panel, the presence of virtual photons in the USC ground state $\ket{E_0}$ allows a current passing only through $\ket{E_0}$ to release photons (the required energy is provided by the energy difference between the reservoirs)~\cite{Cirio2016}.
Figures reproduced with permission from: \textbf{(a)} Ref.~\cite{Lolli2015} \copyright 2015, APS; \textbf{(b)} Ref.~\cite{DeLiberato2009} \copyright 2009 APS; \textbf{(c)} Ref.~\cite{DiStefano2017} \copyright 2017 IOP; \textbf{(d)} Ref.~\cite{Cirio2016} \copyright 2016 APS.
\label{fig:VirtualPhotons}}
\end{figure*}

The photons in the ground state of a system with an atom ultrastrongly coupled to a cavity are not only unable to leave the cavity; they are tightly bound to the atom~\cite{SanchezMunoz2018}. The ground-state photons also cannot be detected by a photoabsorber, even if this absorber is placed inside the cavity, except with very small probability at short times set by the time-energy uncertainty~\cite{DiStefano2017a}. In light of these properties, the ground-state photons in an USC system are considered virtual. However, even though these virtual photons cannot be absorbed by a detector, there are still ways to probe them. One proposal is to measure the change they produce in the Lamb shift of an ancillary probe qubit coupled to the cavity~\cite{Lolli2015} [\figref{fig:VirtualPhotons}(a)]; another proposal is to detect the radiation pressure they give rise to if the cavity is an optomechanical system~\cite{Cirio2017}.

There are also many proposals for how the virtual photons dressing the USC ground state $\ket{E_0}$ (and excited states) can be converted into real ones and extracted from the system. Several of these proposals rely on the rapid modulation of either $g$ or the atomic frequency~\cite{Ciuti2005, DeLiberato2007, Takashima2008, Werlang2008, Dodonov2008, DeLiberato2009, Beaudoin2011, Carusotto2012, Garziano2013, Shapiro2015} [\figref{fig:VirtualPhotons}(b)]. The generation of photons through the modulation of a system parameter in this way requires USC, but not SC, highlighting that $g$ is compared to two different parameters in these regimes~\cite{DeLiberato2017}. A connection can be made between these schemes and the dynamical Casimir effect, where vacuum fluctuations are converted into pairs of real photons when a mirror (or another boundary condition) is moved at high speed~\cite{Moore1970, Johansson2009, Wilson2011, Nation2012, Macri2018}.

Another way to extract virtual photons is to use additional atomic levels. If only the upper transition in a $\Xi$-type three-level atom couples ultrastrongly to the cavity, driving the lower transition can switch that USC on and off~\cite{Gunter2009, Ridolfo2011} to create photons~\cite{Carusotto2012}. The virtual photons in the USC part of such a system can also be released through stimulated emission~\cite{Huang2014}, which opens up interesting prospects for experimental studies of dressed states in the USC regime~\cite{DiStefano2017} [\figref{fig:VirtualPhotons}(c)]. Finally, if the cavity is ultrastrongly coupled to an electronic two-level system, yet another way to release photons from $\ket{E_0}$ is through electroluminescence~\cite{Cirio2016} [\figref{fig:VirtualPhotons}(d)].


\section{Simulating ultrastrong coupling}

\begin{figure*}
\centering
\includegraphics[width=\linewidth]{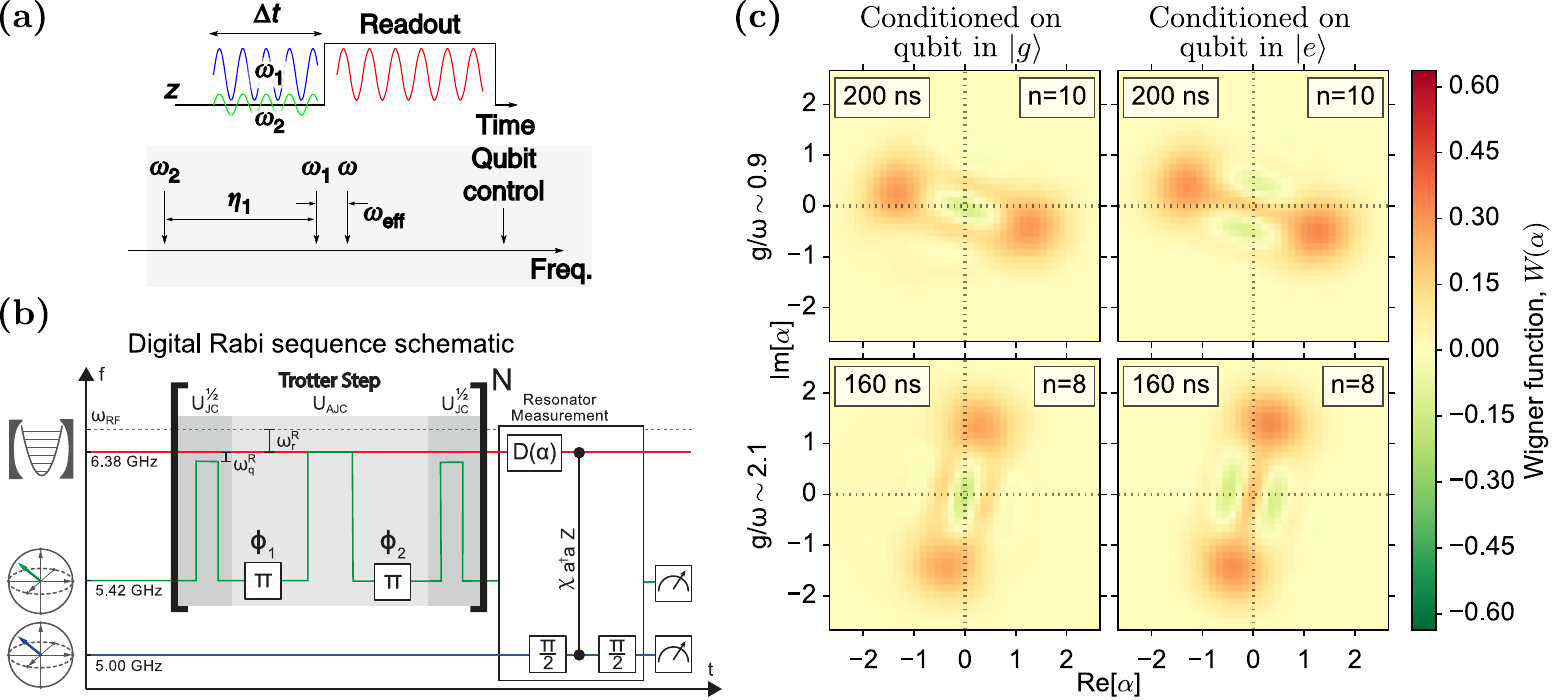}
\caption{Simulations of ultrastrong coupling.
\textbf{(a)} An illustration of the parameters used in the experiment of Ref.~\cite{Braumuller2017} to simulate USC by driving a strongly coupled system with two tones. In the laboratory frame, the qubit has frequency $\epsilon$ and the resonator has frequency $\omega$. Two transversal drives (blue and green curves), with frequencies $\omega_1$ and $\omega_2$, and amplitudes $\eta_1$ and $\eta_2$, respectively, are applied to the qubit. Provided that $\eta_2 \ll \eta_1 = \omega_1 - \omega_2$, the Hamiltonian in the interaction picture has an effective qubit frequency $\epsilon_{\rm eff} = \eta_2 / 2$ and an effective resonator frequency $\omega_{\rm eff} = \omega - \omega_1$, but the effective coupling strength is only halved (from originally being $g$): $g_{\rm eff} = g / 2$. Even if $g \ll \epsilon, \omega$, with the right drive parameters it is possible to simulate $g_{\rm eff} \gtrsim \epsilon_{\rm eff}, \omega_{\rm eff}$. After simulating for a time $\Delta t$, the qubit is detuned for readout.
\textbf{(b)} A sketch of the digital simulation of USC implemented in the experiment of Ref.~\cite{Langford2017}. Here, the bare resonator frequency is $\omega_{\rm r}$ (red line) and the bare qubit frequency is $\omega_{\rm q}$ (green line). In one step of the simulation, the qubit is tuned close to the resonator for a short time, detuned and flipped (marked by $\pi$ and an additional phase shift $\phi_1$), tuned close to resonance again, detuned and flipped (with an additional phase shift $\phi_2$). Due to the two bit flips, the outer interactions with the resonator follow the Jaynes--Cummings Hamiltonian (since the bare coupling $g \ll \omega_{\rm q}, \omega_{\rm r}$; see Box~1) while the middle interaction follow the anti-Jaynes--Cummings Hamiltonian, i.e., the counter-rotating terms. Together, these interactions give the full quantum Rabi Hamiltonian with an effective qubit frequency $\omega_{\rm q}^{\rm R}$ defined by the difference in the detuning of the qubit from the resonator in the different steps, an effective resonator frequency $\omega_{\rm}^{\rm R} = 2 (\omega_{\rm r} - \omega_{\rm RF})$ ($\omega_{\rm RF}$ is a frequency defining the rotating frame, set by $\phi_1 - \phi_2$), and an effective coupling strength $g^{\rm R} = g$. After $N$ repetitions (Trotter steps) of the depicted sequence, the resonator was read out through interaction with another qubit (blue line).
\textbf{(c)} Results from simulations of the USC ground state in Ref.~\cite{Langford2017}. The plots show the Wigner functions for the resonator state conditioned on measuring the qubit being in its ground state (left column) or excited state (right column) for $\eta \sim 0.9$ (upper row, 10 Trotter steps) and $\eta \sim 2.1$ (lower row, 8 Trotter steps). The Schr\"{o}dinger-cat states emblematic of extremely high coupling strengths [compare \figref{fig:EnergyLevelsGroundStateUSC}(e)] are clearly visible.
Figures reproduced with permission from: \textbf{(a)} Ref.~\cite{Braumuller2017} \copyright 2017, NPG; \textbf{(b)}, \textbf{(c)} Ref.~\cite{Langford2017} \copyright 2017, NPG.
\label{fig:SimulatingUSC}}
\end{figure*}

Although the USC regime has been reached in several solid-state systems recently, the experimental effort required to achieve this regime is still considerable. Furthermore, it remains difficult to probe many interesting system properties in these experiments, especially dynamics, for a wide range of parameters. An approach that circumvents these problems is quantum simulation~\cite{Buluta2009, Georgescu2014}, where an easy-to-control quantum system is used to simulate the properties of the quantum model of interest. In 2010, such an approach was used to observe~\cite{Baumann2010} the superradiant phase transition of the Dicke Hamiltonian by placing a Bose-Einstein condensate in an optical cavity and gradually increasing the effective light-matter coupling through an external pump. Another early example is a classical simulation of the dynamics of the parity chains in the quantum Rabi model, realized in an array of femtosecond-laser-written waveguides where the waveguide spacing sets the coupling strength and engineered properties of the waveguides set the effective qubit and resonator frequencies~\cite{Longhi2011, Crespi2012}.
 
Several proposals for quantum simulation of USC rely on driving some part of a strongly coupled system at two frequencies. Then a rotating frame can be found with renormalized parameters, set by the drives, that can be in the USC regime~\cite{Dimer2007, Ballester2012, Grimsmo2013, Pedernales2015, Felicetti2015, Puebla2017, Fedortchenko2017, Aedo2018} (drives can also be used to set effective parameters in other ways~\cite{Felicetti2017, Felicetti2017a}). In 2017, one such proposal~\cite{Ballester2012}, was implemented in a circuit-QED experiment where two drive tones were applied to a superconducting qubit coupled to a transmission-line resonator~\cite{Braumuller2017} [see \figref{fig:SimulatingUSC}(a)]. Starting from a bare $\eta$ below $10^{-3}$, a simulated $\eta$ of above $0.6$ was achieved and the dynamics of population revivals were observed. Recently, the USC was also simulated in a trapped-ion system~\cite{Lv2018} using the proposal of Ref.~\cite{Pedernales2015}, and USC between two resonators was simulated in superconducting circuits~\cite{Markovic2018} following the proposal in Ref.~\cite{Fedortchenko2017}. 

However, external continuous drives are not necessary to define a rotating frame that places the system in the USC regime. An ingenious digital quantum simulation can be realized with a system described by the JC model, as illustrated in \figref{fig:SimulatingUSC}(b). By tuning a qubit in and out of resonance with a resonator, and flipping the qubit in-between, the quantum Rabi Hamiltonian can be simulated~\cite{Mezzacapo2014, Lamata2017} (with multiple qubits, this can be straightforwardly extended to simulating the Dicke Hamiltonian~\cite{Mezzacapo2014, Lamata2017}). This was the approach taken in another recent circuit-QED experiment~\cite{Langford2017}, which simulated $\eta$ up to $1.8$ and observed dynamics in this regime, including the evolution of the photonic Schr\"{o}dinger-cat states in the ground state of the QRM [see \figref{fig:SimulatingUSC}(c)], first predicted in Ref.~\cite{Ashhab2010}. However, it should be noted that the photons in these simulations are always real, not virtual as the photons in a physical USC system are.   


\section{Ultrastrong coupling to a continuum}

\begin{figure*}
\centering
\includegraphics[width=\linewidth]{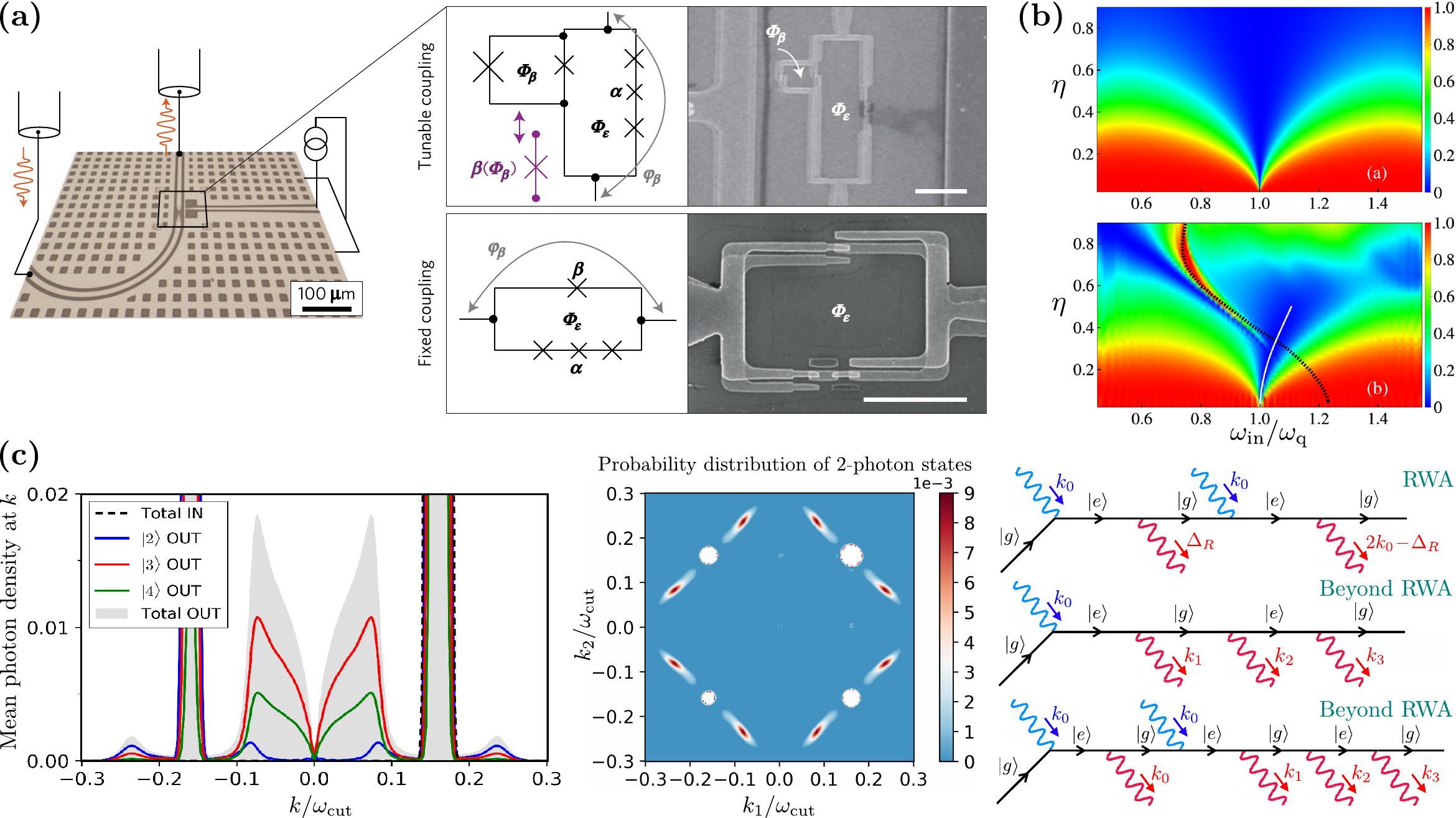}
\caption{Experiment and theory for USC of an atom to an open waveguide.
\textbf{(a)} The experimental setup in Ref.~\cite{Forn-Diaz2017}, the first experiment to achieve USC to an open waveguide. The coplanar waveguide, passing from the input line to the left of the chip to the output line at the top of the chip in the left panel, is interrupted by a loop containing Josephson junctions (crosses in the zoom-in in the right panel). This loop forms the superconducting flux qubit. By adding a second loop, as shown in the upper part of the right panel, the coupling strength can be tuned in situ by an external magnetic flux.
\textbf{(b)} Elastic transmission of a single photon at frequency $\omega_{\rm in}$, travelling in an open waveguide coupled to a qubit with frequency $\omega_{\rm q}$, as a function of normalized coupling strength~\cite{Sanchez-Burillo2014}. The upper panel shows the result using RWA; the lower panel shows the result including the counter-rotating terms in the Hamiltonian. On resonance the single photon is completely reflected by the qubit. The white curve shows the estimated position for the transmittance minimum, which is blue-shifted. The dashed black curve marks the position of a Fano resonance that develops as the coupling increases and the effective qubit frequency is redshifted.
\textbf{(c)} Frequency conversion in off-resonant scattering from a qubit with USC to an open waveguide~\cite{Gheeraert2018}. The left panel shows the scattered photon density at momentum $k$ for a coherent input state at $k_0 = 0.16 \omega_{\rm cut}$ (dashed black curve, mean photon number $\bar n = 0.5$) impinging on a qubit with renormalized frequency $\omega_{\rm q, re} = 0.08 \omega_{\rm cut}$, where $\omega_{\rm cut}$ is the cut-off frequency for the density of states in the waveguide. The blue, red, and green curves show that there are 2-, 3-, and 4-photon states, respectively, in the scattered signal. The middle panel shows the probability distribution of the 2-photon states as a function of the photon momenta $k_1$ and $k_2$. The right panel shows Feynman diagrams demonstrating how the counter-rotating terms in the Hamiltonian (neglected in the RWA) allow more frequency-conversion processes.
Figures reproduced with permission from: \textbf{(a)} Ref.~\cite{Forn-Diaz2017} \copyright 2017 NPG; \textbf{(b)} Ref.~\cite{Sanchez-Burillo2014} \copyright 2014 APS; \textbf{(c)} Ref.~\cite{Gheeraert2018} \copyright 2018.
\label{fig:OpenWaveguideUSC}}
\end{figure*}

An atom can not only couple ultrastrongly to a single harmonic oscillator, but also to a collection or continuum of them. This constitutes an interesting and, so far, less explored regime of the well-known spin-boson model~\cite{Weiss2012}. After USC to a cavity was first realized a decade ago, several theory proposals showed that superconducting circuits was a suitable platform for USC to a continuum (in this case, an open waveguide on a chip)~\cite{Bourassa2009, LeHur2012, Peropadre2013a}. In 2017, such an experiment succeeded~\cite{Forn-Diaz2017} [\figref{fig:OpenWaveguideUSC}(a)] and more demonstrations have followed~\cite{Magazzu2018, PuertasMartinez2018}. Recently, it has also been shown that USC to a continuum could be simulated in superconducting circuits~\cite{Leppakangas2018}, extending the method implemented in Ref.~\cite{Braumuller2017} for simulation of USC to a cavity.

Ultrastrong coupling modifies the physics of an atom in a waveguide dramatically compared to when the coupling is low enough for the RWA to be applicable. Similar to the cavity case, the ground state contains a cloud of virtual photons (in many modes) surrounding the atom~\cite{Peropadre2013a, Snyman2015} and the atom transition frequency experiences a strong Lamb shift~\cite{Weiss2012, Sanchez-Burillo2014, Diaz-Camacho2016}. This considerably changes the transmission of photons in the waveguide; the standard scenario, where the atom reflects single photons on resonance~\cite{Hoi2011}, no longer holds~\cite{Peropadre2013a, Sanchez-Burillo2014, Diaz-Camacho2016, Gheeraert2017} [\figref{fig:OpenWaveguideUSC}(b)]. Instead, similar to the nonlinear-optics-like processes~\cite{Kockum2017a} discussed later in this review, the counter-rotating terms allow various frequency-conversion processes~\cite{Goldstein2013, Sanchez-Burillo2014, Gheeraert2018} [\figref{fig:OpenWaveguideUSC}(c)]. Other new phenomena include decreasing spontaneous emission rate with increasing coupling~\cite{Diaz-Camacho2016}  and spontaneous emission of Schr\"{o}dinger-cat states~\cite{Gheeraert2017}. 


\section{Connections to other models}

The quantum Rabi Hamiltonian (see Box~1) is closely related to a number of other fundamental models and emerging phenomena. These include the Hopfield model (see Box~1), a Jahn-Teller model~\cite{Hines2004, Meaney2010, Larson2008, Bourassa2009, Dereli2012}), a fluctuating-gap model of a disordered Peierls chain~\cite{Levine2004}, as well as renormalization group models, e.g., the spin-boson~\cite{LeHur2012, Leppakangas2018} and Kondo models~\cite{LeHur2012, Goldstein2013, Snyman2015}. The latter two models can be simulated by the superconducting-circuit setups discussed in the previous section [\figref{fig:OpenWaveguideUSC}(a)]. It is counterintuitive, but well-known, that purely electronic phenomena (like the Kondo effect) are closely related to strongly dissipative two-level systems~\cite{Weiss2012}.

Light-matter systems described by a generalized version of the QRM [\eqref{eq:RabiModel} in Box~1 with $g_1\neq g_2$] enable quantum simulations of supersymmetric (SUSY) field theories. Specifically, SUSY can be simulated with coupled resonators, each described by the QRM and tuned to a SUSY point (or line)~\cite{Tomka2015}. The QRM naturally reveals a certain Bose-Fermi duality, which is the central concept of SUSY. This approach enables finding topologically protected subspaces, which may help implementing decoherence-free algorithms for QIP. Moreover, dark matter in cosmology may be explained through SUSY, so superconducting quantum circuits in the USC regime could in principle realize dark-matter simulations on a chip.

The QRM is also equivalent to a Rashba-Dresselhaus model, describing, e.g., a 2DEG with spin-orbit coupling of Rashba and Dresselhaus types interacting with a perpendicular, constant magnetic field~\cite{Tomka2015}. This is a fundamental model of condensed-matter physics, which can be realized in many other systems, e.g., semiconductor heterostructures, quantum wires, quantum dots (confined in parabolic potentials), carbon-based materials, 2D topological insulators, Weyl semimetals, and ultracold neutral atoms.

Furthermore, a superconducting quantum circuit with USC has been suggested for demonstrating vacuum-induced symmetry breaking~\cite{Garziano2014}. This effect is analogous to the Higgs mechanism for the generation of masses of weak-force gauge bosons through gauge-symmetry breaking.


\section{Applications}

\begin{figure*}
\centering
\includegraphics[width=\linewidth]{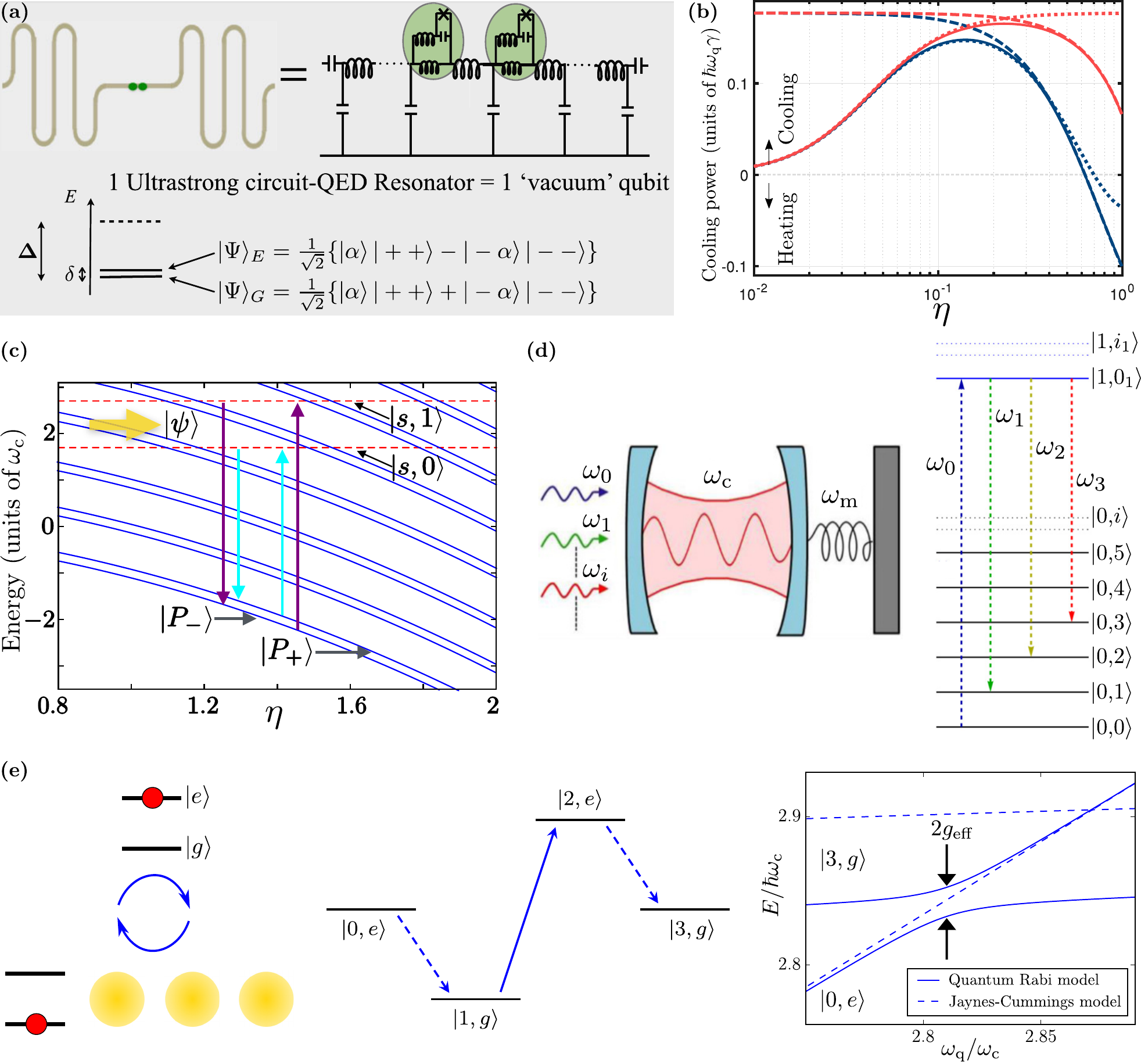}
\caption{Some potential applications of USC.
\textbf{(a)} Protected QIP. In Ref.~\cite{Nataf2011}, it was suggested to use $N$ qubits ($N = 2$ in the figure, which shows superconducting qubits and a transmission-line resonator) ultrastrongly coupled to a resonator to form computational states robust against decoherence. As shown in the lower part of the figure, for high $\eta$ the ground and first excited states of the combined system are entangled coherent resonator states and $\hat \sigma_x$ eigenstates of the qubits [compare \figref{fig:EnergyLevelsGroundStateUSC}(e)]. These two states form a computational subspace that is well separated from other energy levels and protected, to a degree exponentially increasing with $\eta$, from certain decoherence mechanisms. 
\textbf{(b)} Quantum thermodynamics. In Ref.~\cite{Seah2018}, a refrigeration system consisting of three coupled qubits (work, hot, and cold) was studied. The plot shows the cooling power as function of $\eta$. Red curves are results with RWA used for the coupling, while blue curves are results with all coupling terms included; clearly the full coupling is needed to understand the cooling at high $\eta$. Solid curves are calculated using the correct master-equation treatment for USC (see Box~3), while dashed and dotted curves show the result of standard approaches.
\textbf{(c)} Quantum memory. In a setup similar to that of Ref.~\cite{Nataf2011} in panel (a), the two lowest energy levels $\ket{P_{\pm}}$ of a qubit-resonator system in the DSC regime is a good quantum memory~\cite{Stassi2018}. Reading and writing a state $\ket{\psi}$ in the memory is done via an auxiliary atomic level $\ket{s}$ [compare \figref{fig:VirtualPhotons}(c)].
\textbf{(d)} In an optomechanical system (cavity with frequency $\omega_{\rm c}$, moving mechanical mirror with frequency $\omega_{\rm m} \ll  \omega_{\rm c}$), an arbitrary mechanical state can be constructed in a single step when $g \sim \omega_{\rm c}$~\cite{Garziano2015a}. A photonic excitation at frequency $\omega_0$ can be converted into a superposition of mechanical Fock states by simultaneous drives at $\omega_i$ that stimulate transitions.
\textbf{(e)} Nonlinear optics. The left panel shows a schematic depiction of a three-photon Rabi oscillation~\cite{Garziano2015}, where a single two-level atom emits and absorbs three photons. The middle panel shows the virtual transitions in the third-order process that converts the initial state $\ket{i} = \ket{0,e}$ into the final state $\ket{f} = \ket{3, g}$. Dashed arrows are transitions mediated by terms in the JC Hamiltonian and solid arrows are transitions mediated by the counter-rotating terms in the quantum Rabi Hamiltonian. The right panel shows energy levels as a function of $\omega_{\rm q}$. The effective coupling between $\ket{i}$ and $\ket{f}$ on resonance is revealed by the avoided level crossing, which only occurs for the quantum Rabi Hamiltonian (blue solid curves); it is not present in the JC Hamiltonian (blue dashed curves).
Figures reproduced with permission from: \textbf{(a)} Ref.~\cite{Nataf2011} \copyright 2011 APS; \textbf{(b)} Ref.~\cite{Seah2018} \copyright 2018 APS; \textbf{(c)} Ref.~\cite{Stassi2018} \copyright 2018 APS; \textbf{(d)} Ref.~\cite{Garziano2015a} \copyright 2015 APS.
\label{fig:Applications}}
\end{figure*}

Why do we need USC when we already have SC? The simplest answer is that USC enables more efficient interactions. For example, the coupling between a single photon and a single emitter results in significant nonlinearity, which has been used in electro-optical devices operating in the SC regime. Increasing $\eta$ from SC to USC results in better performance of such devices, e.g., faster control and response even for shorter lifetime of the device components. Some quantum effects (including quantum gates) in specific realistic short-lifetime systems cannot be observed below USC.

The list of emerging applications of USC goes on much longer: QIP, quantum metrology, nonlinear optics, quantum optomechanics, quantum plasmonics, superconductivity, metamaterials, quantum field theory, quantum thermodynamics, and even chemistry QED and materials science. Below, we discuss some of these applications in greater detail.

Another question arises: can one predict and observe entirely new phenomena in the USC or DSC regimes? A simple example is the experimental observation of new stable states of matter, i.e., entangled hybrid light-matter ground states in the DSC regime~\cite{Yoshihara2017} [see also \figref{fig:EnergyLevelsGroundStateUSC}].


\subsection{Quantum information processing}

Cavity- and circuit-QED systems in the USC regime are especially useful for quantum technologies like quantum metrology (e.g., novel high-resolution spectroscopy~\cite{Ruggenthaler2018} utilizing smaller linewidths and improved signal-to-noise ratio) and QIP. For QIP, coherent transfer of excitations between light and matter is particularly important. Such transfer can be achieved in the SC regime, but it can be much more efficient in the USC regime. Other QIP applications of USC include (i) extremely fast quantum gate operations~\cite{Romero2012, Wang2017}, (ii) efficient realizations of quantum error correction~\cite{Stassi2017}, (iii) quantum memories~\cite{Kyaw2015, Stassi2018} [\figref{fig:Applications}(c)], (iv) protected QIP~\cite{Nataf2011} [\figref{fig:Applications}(a)], and (v) holonomic QIP~\cite{Wang2016c}. The advantages are not only shorter operation times, but also simpler protocols, where the natural evolution of a USC system replaces a sequence of quantum gates~\cite{Stassi2017}. Some of these proposals also exploit the entangled ground states and parity symmetry.


\subsection{Modifying standard quantum phenomena}

Increasing $\eta$ from SC to USC, various standard quantum phenomena are often changed drastically. Examples include the Purcell effect~\cite{DeLiberato2014}, electromagnetically induced transparency and photon blockade~\cite{Ridolfo2012, LeBoite2016}, spontaneous emission spectra~\cite{Cao2011}, the Zeno effect~\cite{Lizuain2010, Cao2012}, refrigeration in quantum thermodynamics~\cite{Seah2018} [\figref{fig:Applications}(b)], and photon transfer in coupled cavities~\cite{Felicetti2014}. Such modified effects offer new emerging applications. In particular, light-induced topology~\cite{Lindner2011, Claassen2017, Hubener2017} and quantum plasmonics~\cite{Tame2013} with SC can, in principle, be improved and diversified with USC. Another intriguing development is that USC may help understanding unconventional superconductivity through studies of light-enhanced (i.e., polaritonically-enhanced)~\cite{Sentef2018} and photon-mediated~\cite{Schlawin2018} superconductivity.


\subsection{Higher-order processes and nonlinear optics}

The inclusion of the counter-rotating terms in the quantum Rabi Hamiltonian also allows predicting higher-order processes. A prominent example is deterministic nonlinear optics (or vacuum-boosted nonlinear optics) with two-level atoms and (mostly) virtual photons in resonator modes~\cite{Kockum2017a, Stassi2017, Kockum2017}. These implementations, in contrast to conventional realizations of various multi-wave mixing processes in nonlinear optics, can reach perfect efficiency, need only a minimal number of photons, and require only two atomic levels. The counter-rotating terms can also be leveraged in USC optomechanics to rapidly construct mechanical quantum states~\cite{Garziano2015a} [\figref{fig:Applications}(d)] or observe the dynamical Casimir effect~\cite{Macri2018}.

Many nonlinear-optics processes can be described in terms of higher-order perturbation theory involving virtual transitions, where the system passes from an initial state $\ket{i}$ to the final state $\ket{f}$ via a number of virtual transitions to intermediate states. These virtual transitions need not conserve energy, but their sum, the transition from $\ket{i}$ to $\ket{f}$, does. When the light-matter coupling strength increases, the vacuum fluctuations of the electromagnetic field become able to induce such virtual transitions, replacing the role of the intense applied fields in nonlinear optics. In this way, higher-order processes involving counter-rotating terms can create an effective coupling between two states of the system ($\ket{i}$ and $\ket{f}$) with different number of excitations~\cite{Kockum2017a}. The strength of the effective coupling $g_{\rm eff}$ approximately scales as $g \eta^{n}$, where $n$ is the number of intermediate virtual states visited by the system on the way between $\ket{i}$ and $\ket{f}$ [an $(n+1)$th-order process].

If the light-matter coupling is sufficiently strong, $g_{\rm eff}$ becomes larger than the relevant decoherence rates in the system (i.e., the \textit{effective} coupling can be termed \textit{strong}). In this case, the resulting coupling is deterministic, a highly desirable feature for practical applications in quantum technologies. Such a resonant coupling between two states with different number of excitations was observed in one of the first USC experiments in 2010~\cite{Niemczyk2010}. 

Subsequent theoretical investigations have shown that these higher-order processes can give rise to intriguing novel CQED effects, e.g., anomalous quantum Rabi oscillations, where a two-level atomic transition can coherently emit or absorb photon pairs or triplets~\cite{Ma2015, Garziano2015} [see \figref{fig:Applications}(e)], or multiple atoms jointly absorb or emit a single photon, each atom taking or providing part of the photon energy~\cite{Garziano2016}. These novel deterministic processes enrich the possibilities of using cavity-QED for the development of efficient protocols for quantum technologies.

As discussed in the preceding section, superconducting quantum circuits with USC can also be used to simulate other fundamental models and testing their predictions, e.g., in quantum field theory and solid-state physics.  We believe that these connections of the QRM to other fundamental models in various branches of physics can mutually stimulate research in all these fields by finding new analogs of condensed-matter effects in quantum-optical systems and vice versa.


\subsection{Chemistry with ultrastrong coupling}

There is increasing interest, theoretical and experimental, in the study of SC and USC CQED with molecular ensembles. This may lead to new routes to control chemical bonds and reactions (e.g., dynamics, kinetics, and thermodynamics) at the nanoscale level. Such photochemistry of molecular polaritons in optical cavities~\cite{Galego2015, Ebbesen2016, Bennett2016} is sometimes referred to as cavity~\cite{Kowalewski2016} (or cavity-controlled~\cite{Herrera2016}) chemistry, polariton chemistry~\cite{Martinez-Martinez2018, Sentef2018}, or QED chemistry~\cite{Ruggenthaler2014, Schafer2018}. This interest was partially triggered by the experiment of Ref.~\cite{Schwartz2011}, which demonstrated the control of the coupling (from WC to USC) between photochromic spiropyran molecules and light in a low-Q metallic cavity. In this and related experiments~\cite{Ebbesen2016}, USC was reached by \textit{collective} coupling of many molecules to the cavity mode. To achieve USC (or even only SC) for a \textit{single} molecule is much more demanding~\cite{Chikkaraddy2016}.

Recent studies show that the \textit{excited}-state reactivity of photochemical processes (like catalysis and photosynthesis) in molecules in nanocavities can be substantially modified by SC and USC~\cite{Galego2015, Herrera2016, Cwik2016, Martinez-Martinez2018}. The reason is that $g$ is comparable with the energies of vibrational and electronic transitions in molecules and their coupling~\cite{Galego2015}. In particular, a better control of chemical reactions can be realized via polaron decoupling, induced by SC or USC, of electronic and nuclear degrees of freedom in a molecular ensemble~\cite{Herrera2016}. Possibilities and limitations of applying USC to change the electronic \textit{ground} state of a molecular ensemble to control chemical reactions have also been investigated~\cite{Galego2015, Martinez-Martinez2018}. It was shown that some molecular observables depend solely on single-molecule couplings, while others (e.g., related to electronically excited states) also can be modified by collective couplings. Moreover, low-barrier chemical reactions can be affected by the quantum interference of different reaction pathways occurring simultaneously in multiple molecules ultrastrongly coupled to a cavity~\cite{Martinez-Martinez2018}.

Some of these works~\cite{Bennett2016, Kowalewski2016} were based on the QRM as in the standard quantum-optical approach. The Dicke model with antiresonant terms (see Table~B1.I) was applied to study many molecules coupled to a surface plasmon~\cite{Martinez-Martinez2018}. Some other works~\cite{Schafer2018} used a powerful QED density-functional formalism of QED chemistry~\cite{Ruggenthaler2014}. This formalism unifies quantum optics and electronic-structure theories by treating a QED system composed of matter and light as a quantum liquid. The original formalism works well for SC, but becomes much less efficient (``extensively cumbersome'') in the USC regime~\cite{Schafer2018}.


\section{Conclusion and outlook}

As we described in this review, many intriguing physical effects have already been predicted in the USC regime of light-matter interaction. However, related experiments have been limited to increasing the light-matter coupling strength and verifying it by standard transmission measurements. Now that USC has been reached in a broad range of systems, we believe that it is high time to explore experimentally the new interesting phenomena specific to USC and, finally, to find their useful applications. A few decades ago, CQED in the WC and SC regimes was following the same route, which lead to important applications in modern quantum technologies. We therefore believe that USC applications have the potential to make a profound impact.


\section{Acknowledgements}

Numerical simulations were performed using QuTiP~\cite{Johansson2013}. We thank Yu-xi Liu, Xiu Gu, Omar Di Stefano, and Alessandro Ridolfo for useful discussions. We also thank Alessio Settineri, Mauro Cirio, and Shahnawaz Ahmed for technical assistance with some of the figures. A.F.K.~acknowledges partial support from a JSPS Postdoctoral Fellowship for Overseas Researchers (P15750). A.M.~and F.N.~acknowledge support from the Sir John Templeton Foundation. S.D.L.~acknowledges support from a Royal Society research fellowship. F.N.~also acknowledges support from the MURI Center for Dynamic Magneto-Optics via the Air Force Office of Scientific Research (AFOSR) award No.~FA9550-14-1-0040, the Army Research Office (ARO) under grant No.~73315PH, the Asian Office of Aerospace Research and Development (AOARD) grant No.~FA2386-18-1-4045, the Japan Science and Technology Agency (JST) through the ImPACT program and CREST Grant No.~JPMJCR1676, the Japan Society for the Promotion of Science (JSPS) through the JSPS-RFBR grant No.~17-52-50023, and the RIKEN-AIST Challenge Research Fund.

\bibliography{ReferencesUSC}

\end{document}